\documentclass[fleqn,usenatbib]{mnras}

\usepackage{subcaption}

\DeclareRobustCommand{\VAN}[3]{#2}
\let\VANthebibliography\thebibliography
\def\thebibliography{\DeclareRobustCommand{\VAN}[3]{##3}\VANthebibliography}

\usepackage{graphicx}	
\usepackage{amsmath}	
\usepackage{amssymb}	
\usepackage{bm}

\title[Synthetic polarization maps for a Local Bubble-like cavity]{Synthetic dust polarization emission maps at 353\,GHz for an observer placed inside a Local Bubble-like cavity}

\author[E. Maconi et al.]{E. Maconi$^{1,2,3}$\thanks{E-mail: e.maconi1@campus.unimib.it},
J. D. Soler$^{4}$\thanks{E-mail: juandiegosolerp@gmail.com},
S. Reissl$^{1}$,
P. Girichidis$^{1}$,
R. S. Klessen$^{1,5}$,
P. Hennebelle$^{6}$,
S. Molinari$^{4}$,
\newauthor
L. Testi$^{7}$,
R. J. Smith$^{8}$,
M. C. Sormani$^{1}$,
J. W. Teh$^{1}$,
A. Traficante$^{4}$.
\\
$^{1}$Universit\"{a}t Heidelberg, Zentrum f\"{u}r Astronomie, Institut f\"{u}r Theoretische Astrophysik, Albert-Ueberle-Str. 2, 69120, Heidelberg, Germany.\\
$^{2}$Università degli Studi di Milano-Bicocca. Physics Department, p.zza della Scienza, 3, 20126. Milan, Italy.\\
$^{3}$ Universit\"{a}t Wien, Institut f\"{u}r Astrophysik, T\"{u}rkenschanzstrasse 17, 1180 Wien, Austria.\\
$^{4}$Istituto di Astrofisica e Planetologia Spaziale. INAF. Via Fosso del Cavaliere, 100. 00133. Roma, Italy.\\
$^{5}$Universit\"{a}t Heidelberg, Interdisziplin\"{a}res Zentrum f\"{u}r Wissenschaftliches Rechnen, Im Neuenheimer Feld 205, D-69120 Heidelberg, Germany. \\
$^{6}$AIM, CEA, CNRS, Université Paris-Saclay, Université Paris Diderot, Sorbonne Paris Cité, F-91191 Gif-sur-Yvette, France.\\
$^{7}$Dipartimento di Fisica e Astronomia "Augusto Righi". Viale Berti Pichat 6/2, Bologna, Italy.\\
$^{8}$Jodrell Bank Centre for Astrophysics, Department of Physics and Astronomy, University of Manchester, Oxford Road, Manchester M13 9PL, UK.\\
}

\date{Accepted XXX. Received YYY; in original form ZZZ}

\pubyear{2023}

\begin{document}
\label{firstpage}
\pagerange{\pageref{firstpage}--\pageref{lastpage}}
\maketitle

\begin{abstract}
We present a study of synthetic observations of polarized dust emission at 353\,GHz as seen by an observer within a cavity in the interstellar medium (ISM). The cavity is selected from a magnetohydrodynamic simulation of the local ISM with time-dependent chemistry, star formation, and stellar feedback in form of supernova explosions with physical properties comparable to the Local Bubble ones. 
We find that the local density enhancement together with the coherent magnetic field in the cavity walls makes the selected candidate a translucent polarization filter to the emission coming from beyond its domains. This underlines the importance of studying the Local Bubble in further detail. The magnetic field lines inferred from synthetic dust polarization data are qualitatively in agreement with the all-sky maps of polarized emission at 353\,GHz from the {\it Planck} satellite in the latitudes interval $15^{\circ} \lesssim |b| \lesssim 65^{\circ}$. As our numerical simulation allows us to track the Galactic midplane only out to distances of $250\,$pc, we exclude the region $|b| \lesssim 15^{\circ}$ from our analysis. At large Galactic latitudes, our model exhibits a high degree of small-scale structures. On the contrary, the observed polarization pattern around the Galactic poles is relatively coherent and regular, and we argue that the global toroidal magnetic field of the Milky Way is important for explaining the data at $|b| \gtrsim 65^{\circ} $. We show that from our synthetic polarization maps, it is difficult to distinguish between an open and a closed Galactic cap using the inferred magnetic ﬁeld morphology alone.
\end{abstract}



\section{Introduction}

Astronomical observations of interstellar reddening and  soft-diffuse X-rays indicate that the Sun lies within a cavity of low-density, high-temperature and warm plasma surrounded by a shell of cold, neutral gas and dust \citep[see e.g.][]{sanders1977,lucke1978,coxANDreynolds1987, Snowden2015b,Linsky2021,ZuckerEtAl2022}. 
The edges of this cavity, commonly called the ``Local Bubble'', are producing a portion of the signal in the observations of dust-polarized thermal emission used in the study of the Galactic magnetic field and the foreground to cosmic microwave background (CMB) radiation \citep{PlanckXLII2016,PlanckXI2018}.
Therefore, understanding the polarization signal produced by the Local Bubble is of crucial importance for Galactic astronomy and cosmology.

Starlight polarization at optical-to-infrared wavelengths and dust polarized emission most likely arise from asymmetric dust grains that are aligned perpendicular to the magnetic field lines \citep[see][for a review]{andersson2015}.
For a long time, extended polarization maps were restricted to observations of the stellar polarization data available for a discrete set of lines-of-sight \citep[LOSs;][]{heiles2000}.
The convergence of the starlight polarization orientations toward the presumed tangent points of the local spiral arm motivated the modelling of the polarized signal as the result of the curvature of the Galactic magnetic field \citep{heiles1996}.
In that model, the local magnetic field closely follows the curvature of the spiral arm, as observed in other galaxies \citep{beck2013,BorlaffEtAl2021}, and the convergence of the field lines is an effect of our perspective as observers within the arm.  

The European Space Agency's {\it Planck} satellite produced the first all-sky map of the polarized emission from dust at sub-millimetre wavelengths \citep{PlanckI2016}.
The {\it Planck} map of the polarized emission at 353\,GHz revealed a regular pattern of magnetic field orientations toward the Galactic caps \citep{PlanckXLIV2016}.
Inspired by the regular pattern produced by the expansion of a super-bubble in a uniform magnetic field, researchers have modelled the {\it Planck} observations  as the result of a magnetic field frozen into the walls of an inclined spheroidal bubble \citep{FerriereEtAl1991,alves2018,PelgrimsEtAl2020}.

In this paper, we present a study of the polarization imprint produced by a Local Bubble-like cavity identified in magnetohydrodynamic (MHD) simulations of a multi-phase stratified medium with overall properties selected to match the Milky Way at the solar radius. The numerical simulations are part of the SILCC (SImulating the Life-Cycle of molecular Clouds) project \citep{WalchEtAl2015,GirichidisEtAl2016b}.  Once the candidate cavity has been selected, the radiative transfer (RT) code {\tt POLARIS} \citep{Reissl2016,Reissl2019} is used to produce synthetic dust polarization emission maps at 353\,GHz for an observer placed inside the bubble analogue. Thanks to the  RT simulations, the 3D information remains accessible along each LOS and the origin of the polarized signal can be explored.

This paper is organized as follows.
In Section~\ref{sec:localbubble} the known properties of the Local Bubble are reviewed.
Section~\ref{sec:sims} presents the MHD numerical simulations, the bubble selection process, and the method used to identify the cavity walls.
Section~\ref{sec:synthobs} describes the post-processing setup used to produce the synthetic observations.
The results are discussed in Section~\ref{sec:discussion} and the conclusions are presented in Section~\ref{sec:conclusions}.

\section{Properties of the Local Bubble }\label{sec:localbubble}

\begin{table*}
	\centering
	\caption{Key properties of the Local Bubble.}
	\label{tab:localbubble}
	\begin{tabular}{lcll} 
		\hline
		Quantity & & Value & Reference\\
		\hline
		Shell total mass & $M_{\rm{shell}}$ & $1.4 \times 10^6\,\mathrm{M}_{\odot}$  & \cite{ZuckerEtAl2022}\\
		Shell expansion vel. &$v_{\rm{shell}}$ & $6.7  \, \rm{km} \, \rm{s}^{-1}$  & \cite{ZuckerEtAl2022}\\
		Thermal pressure & $P/k_{\rm B}$ & 10,700\,cm$^{-3}$\,K & \cite{SnowdenEtAl2014}\\
		Mean field strength &$\left<\left|B\right|\right>$ & 0.5 to 2\,$\mu$G & \cite{XuHan2019} \\
		\hline
	\end{tabular}
\end{table*}

The Local Bubble is thought to be the product of star formation and supernova (SN) explosions that occurred over the past 10-15 Myr, as recently reconstructed by \cite{ZuckerEtAl2022}. 
The modelling of the cavity expansion and the tracing back of the solar orbit indicates that the Sun entered the Local Bubble approximately 5\,Myr ago.
Additional SN explosions have been linked to anomalous concentrations of the radionuclide $^{60}$Fe in deep-sea  layers dating from about 2.2\,Myr ago \citep{BreitschwerdtEtAl2016,SchulreichEtAl2018}. The mass of gas swept away by the SN explosions, and which now forms the shell, has been estimated in $M_{\rm{shell}}=(1.4\pm0.6) \times 10^6\,\mathrm{M}_{\odot}$ and this shell is still expanding with a velocity of $v_{\rm{shell}}=(6.7 \pm 0.4) \, \rm{km} \, \rm{s}^{-1}$ \citep{ZuckerEtAl2022}.

The high-temperature plasma enclosed by the Local Bubble  shell was identified as the source of the  unexpectedly intense diffuse flux of $0.25\,\rm{keV}$ X-rays observed over the whole sky by the ROSAT satellite \citep{SnowdenEtAl1997,GaleazziEtAl2014,puspitarini2014,Snowden2015b}. 
The nature of the inner environment of the  Local Bubble, whether it consists of a low-density million-degree plasma ($10^6$\,K) heated by supernova shocks or a combination of hot and warm (10,000–20,000 K) components, is still a subject of ongoing debate \citep[see e.g.][]{Linsky2021}.

The Local Bubble was identified as the origin of the high-latitude radio emission measured by the ARCADE 2 balloon-borne telescope and other instruments at frequencies below about 3\,GHz.
\citep{SunEtAl2008,krauseANDhardcastle2021}.
More recently, the dusty boundary of the Local Bubble was mapped using the extinction in the emission from stars with known distances \citep{LallementEtAl2014,ZuckerEtAl2022}.

Table~\ref{tab:localbubble} summarizes the main Local Bubble properties, which are derived as follows.
The thermal pressure of the plasma in the Local Bubble $P/k_{B}$\,$=$\,10,700\,cm$^{-3}$\,K is derived from the combination of the magnetic field measurement outside of the heliosphere by the Voyager~1 spacecraft and the soft X-rays observations from sounding rockets and the ROSAT satellite \citep{SnowdenEtAl2014}.
The X-ray emission leads to a thermal electron density $n_{e,\mathrm{X}}$\,$=$\,$(4.68$\,$\pm$\,$0.47)$\,$\times$\,$10^{-3}$\,cm$^{-3}$, which is typical for superbubbles, including X-ray bright ones, as shown in 3D numerical simulations \citep[see e.g.][]{KrauseEtAl2013a,KrauseEtAl2013b}.

Using the Faraday rotation measure (RM) of 494 pulsars within 3\,kpc from the Sun, \cite{XuHan2019} estimated that the Local Bubble edge is located at between 80 and 140\,pc from the Sun. 
These observations indicate that the Local Bubble's mean magnetic field strength is between 0.5 and 2.0\,$\mu$G.
For eight pulsars close to the edge of the Local Bubble, \citet{XuHan2019} report a mean dispersion measure (DM) of 42\,cm$^{-3}$\,pc with a standard deviation of 20\,cm$^{-3}$\,pc, which corresponds to a column of free, thermal electrons $N_{\mathrm{e}}$\,$=$\,$(1.3$\,$\pm$\,$0.6)$\,$\times$\, 10$^{20}$\,cm$^{-2}$.

The column density of free electrons and the RM are most likely produced by the bubble wall, in the ionized mixing layer between the bubble interior and the cold shell \citep{XuHan2019}.
Warm clouds within the Local Bubble have characteristic sizes of several parsecs and electron densities
of the order of $n_{\mathrm{e,WC}}$\,$=$\,$10^{-1}$\,cm$^{-3}$ \citep[see e.g.][]{gry2017,linsky2019}.
Assuming a total warm cloud path length of 10\,pc, the corresponding free electron column is $N_{\mathrm{e,wc}}$\,$=$\,$3\times 10^{18}\,\mathrm{cm}^{-2}\,$.
Hence, neither the hot X-ray plasma nor the warm clouds and filaments contribute significantly to the pulsar DMs. 

\section{Simulated Local Bubble}\label{sec:sims}

\subsection{Simulation and selection of the bubble} \label{subSec:simul+select-bubble}
We identify a Local Bubble candidate in the set of numerical simulations presented in \citet{GirichidisEtAl2018b} and \cite{girichidis2021}, which are part of the  SILCC project \citep{WalchEtAl2015,GirichidisEtAl2016b}.
These simulations consider a multiphase medium in a domain of a 500\,$\times$\,500\,$\times$\,500\,pc$^{3}$  with conditions close to the local neighbourhood in terms of the total gas surface density ($\langle\Sigma_\mathrm{gas}\rangle=10\,\mathrm{M}_\odot\,\mathrm{pc}^{-2}$), the velocity dispersion ($\langle v_\mathrm{gas}\rangle\sim10\,\mathrm{km\,s^{-1}}$), the average magnetization ($B \sim 0.1-10\,\mu\mathrm{G}$) as well as the chemical composition and the local rate of SN explosions \citep[see][]{GirichidisEtAl2018b}.
The simulations run with an adaptive mesh refinement with the highest resolution set to $0.98\,\mathrm{pc}$.

The initial gas distribution is stratified following a Gaussian profile with a scale height of $60\,\mathrm{pc}$, aiming to reproduce the conditions in a portion of the Galactic plane approximately at the solar radius. 
The gas is initially at rest and in pressure equilibrium.
The initial magnetic field is set along the $x$ direction and has a central intensity of $B(z=0)$\,$=$\,3\,$\mu$G, that scales in the vertical direction as ${ B(z) = B(z=0)(\rho(z)/\rho(z=0))^{1/2} }$.

The simulations include SN feedback, which stirs the medium and creates a three-phase medium comparable to the observed local interstellar medium (ISM).
The individual and clustered SN explosions create dense filaments and clouds of cold gas, mainly in the regions in which shells collide, and hot voids, primarily close to the local explosion sites of clustered SN. 

We collect a series of snapshots from the simulations and proceed by selecting a set of bubbles by eye in each time step.
For each bubble we compute the average density, the total mass in ionized gas, and the mass- and volume-weighted magnetic field and temperature in order to identify the cavity whose properties best match the Local Bubble.
Our candidate bubble results from the explosion of 17 clustered SN explosions, in agreement with the 14-20 SN events that appear to be responsible for the Local Bubble \citep{Fuchs2006}. In our simulation, the gas mass swept away by the SN explosions is $M_{\rm{shell}} \, \sim \, 2 \times 10^5\,\mathrm{M}_{\odot}$, an order of magnitude lower compared to \cite{ZuckerEtAl2022}. However, we would like to emphasize that the total energy driving the expansion of the cavity is more important than the shell mass. For one thing, the measured mass of the shell around the bubble depends on the integration length and the structure of the surrounding medium. In particular, if the vicinity of the bubble is highly structured a significant fraction of the shell mass is located in small concentrated clouds. In fact, in the Paper of \citet{ZuckerEtAl2022}  the estimate of the shell mass around the Local Bubble includes the nearby clouds such as Taurus and Ophiuchus (private communication), which we do not have in our simulation. Consequently, the amount of locally concentrated mass only occupies a very small solid angle and the analysis of the full sky around the Local Bubble is only marginally affected by such high density nearby clouds.
Moreover, we aim to select a cavity that best fits the properties of the Local Bubble, but we are aware of the impossibility of finding a perfect candidate.
For more information on the simulations setup and on the selection of the candidate cavity, we refer the reader to Appendix~\ref{Appendix:sims} and to \cite{girichidis2021}. 

\begin{figure*}
    \centering
    \includegraphics[width=0.95\textwidth]{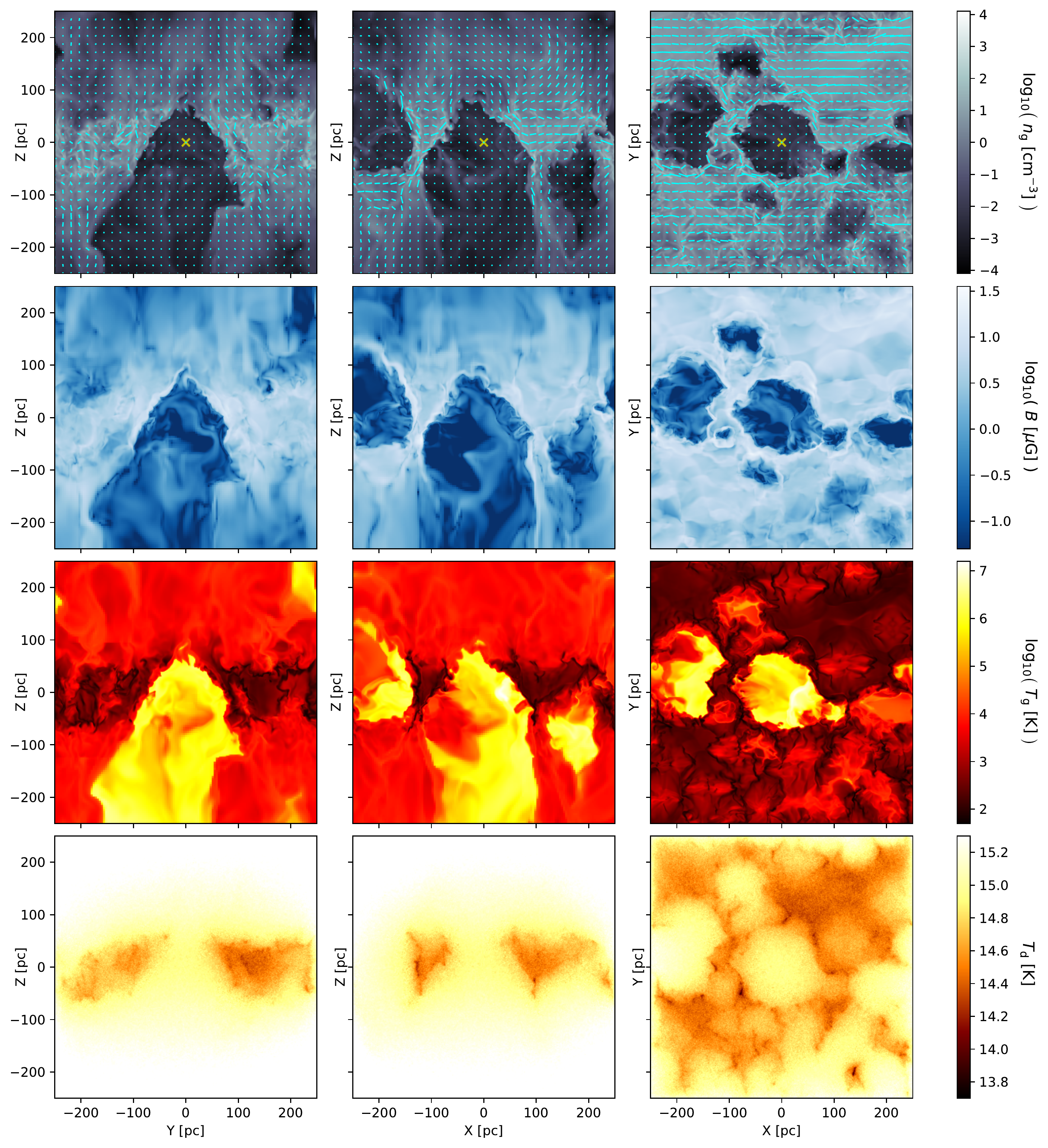}
    
    \caption{From top to bottom: gas number density ($n_{\mathrm{g}}$), total magnetic field strength ($B$), gas temperature ($T_{\mathrm{g}}$), and dust temperature ($T_{\mathrm{d}}$) in 1-pc-thick cuts of the simulation domain centred on the Local Bubble candidate.
    The top panel shows the orientation of the magnetic field, in cyan pseudo-vectors normalized to $|\textbf{B}_{\rm max}|$; the position of the observer used in the synthetic observations is shown as a yellow cross.
    }
    \label{fig:allCuts-lowRes}
\end{figure*}

Fig.~\ref{fig:allCuts-lowRes} shows the gas density, the gas temperature, and the total magnetic field strength for the candidate bubble in cuts through the centre. 
The bottom panel of Fig.~\ref{fig:allCuts-lowRes} also reports the dust temperature computed by {\tt POLARIS}, as further discussed in Sect.~\ref{sec:synthobs}.
The figure illustrates that our cavity is not isolated but surrounded by other expanding bubbles, which influence each other as they expand. This picture is in agreement with the observed structure of the local ISM, where the Local Bubble coexists with other SN-blown cavities \citep[see e.g.][]{Soler2018, Bracco2020, Bialy2021, ZuckerEtAl2022}.
Moreover, since the selected cavity has an open and a closed cap, we can study the differences in polarization properties and column density between the two configurations; this has observational value because evidence suggests that the local interstellar cavity could be open in the galactic halo, forming the so-called ``Local Chimney'' \citep[][]{Cox1974, Welsh1999, Lallement2003,ZuckerEtAl2022}.

\subsection{Defining the walls of the simulated bubble}\label{subSect:methodWalls}

Once we identify a possible Local Bubble candidate  in the simulations, we determine the inner surface of the cavity using the column density maps provided by {\tt POLARIS}.
{\tt POLARIS} allows an angular sampling of the celestial sphere according to a {\tt HEALPix} pixelization \citep{Gorski2005}. We measure the emerging polarization signal and the column density values on 70 concentric spheres centred on the cavity and with an increasing radius step of $5\,$pc. Each sphere is divided into 786432 pixels, corresponding to the {\tt HEALPix} parameter $N_{\rm side}=256$ and to an angular resolution of $13'.7$. 

For each LOS, we construct a radial profile of the column density $N_{\rm H}(r)$, with $r$ being the distance from the cavity centre i.e. the observer.
We define the radius of the Local Bubble inner surface ($r_{\rm in}$) as the distance from the cavity centre, starting from the observer and going outward, at the location where $N_{\rm H}$ has its first relative increase greater than 0.9\,\%.
We choose this value empirically after several attempts.  
We avoid the selection of isolated dense structures within the cavity by setting a threshold on temperature and  magnetic field strength. If the temperature and magnetic field at the selected point are greater than $10^{4.5}\,\mathrm{K}$ and less than $10^{-1.2}\,\mu\mathrm{G}$, it is not registered as part of the cavity wall.

The outer surface of the Local Bubble candidate is established following the procedure defined in \cite{PelgrimsEtAl2020}, where the geometry of the Local Bubble shell is determined using the relative values of the observed dust density. In their work, they draw LOSs starting from the Sun and going outward, with a radial sampling step of 2.5\,pc, and, for each LOS, they compute the differential extinction $A_{\rm{V}}^{'}(r)$. The inner surface of the Local Bubble, $r_{\rm{inner}}$, is then defined as the distance to the first point where the curve has a change from convex to concave. Similarly, the outer surface of the cavity is defined as the point, after  $r_{\rm{inner}}$, where $A_{\rm{V}}^{'} (r)$ changes from concave to convex.

\begin{figure}

    \centering
    \includegraphics[width=0.48\textwidth]{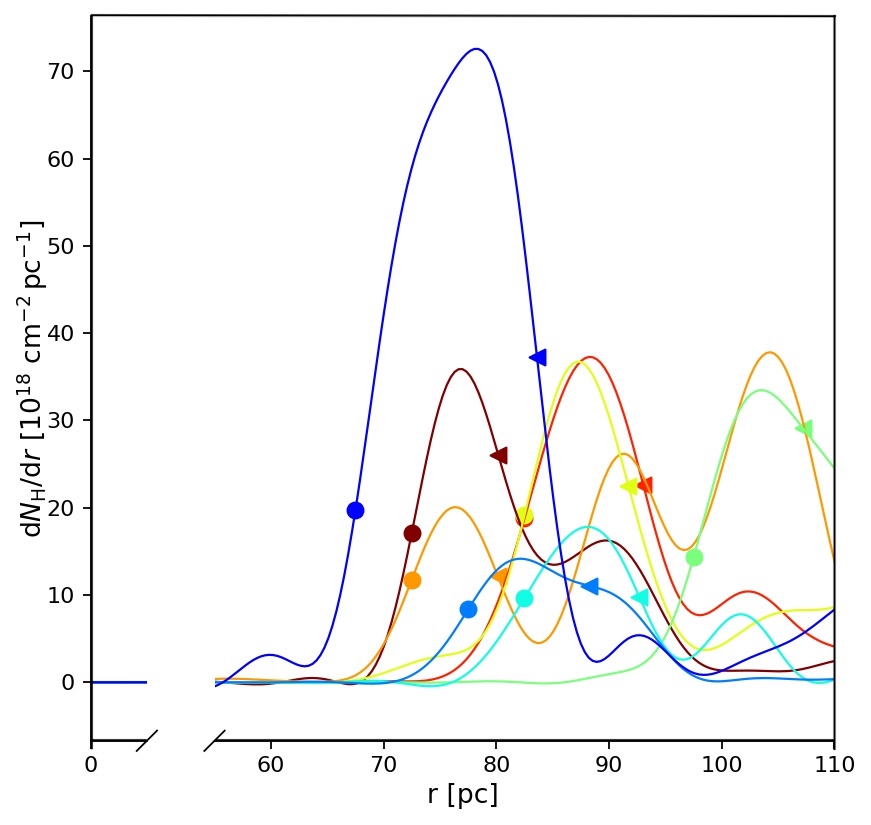}
    
    \caption{Radial profiles of the differential column density $\mathrm{d}N_{\mathrm{H}}\mathrm{(}r\mathrm{)}/\mathrm{d}r$ are shown for eight LOSs in the $x-y$ plane equally spaced by $45^\circ$. The eight LOSs are presented in the upper left panel of Fig.~\ref{fig:polRays_XYplane-lowRes}. The inner and outer boundaries of the selected cavity, as computed in Section~\ref{subSect:methodWalls}, are indicated on each curve with a filled circle and triangle, respectively.}
    \label{fig:differential_NH}

\end{figure}

We use the \cite{PelgrimsEtAl2020} method to determine the outer shell of our Local Bubble candidate, giving as input the inner surface radius previously computed.
We smooth the $N_{\rm H}$ {\tt HEALPix} maps with a Gaussian symmetric beam ($\sigma$\,$=$\,0.015\,rad) and we compute, for each LOS, the differential of the column density $\Delta N_{\rm H}(r)$ and its first and second derivatives with respect to the radius.
The outer surface radius ($r_{\rm{out}}$) is set to be the distance where the curves have their first inflection point beyond the inner wall. 
Due to the complexity of gas distribution, this method to find the Local Bubble walls may fail in some directions but we can confirm this is only the case for a small percentage of LOSs. We show the differential column density curves for eight LOSs selected in the $x-y$ plane in Fig.~\ref{fig:differential_NH}.
We calculate the cavity thickness as the mean value of $r_{\rm{out}} - r_{\rm{in}}$, which for our cavity results in $\sim$\,$14\,$pc.
The {\tt HEALPix} spherical-coordinate information, which describes the computed distance to the Local Bubble walls along every LOS, is later used in this paper to select the part of the bubble enclosed by the outer wall and to study its influence on the polarization of light.

\begin{figure*}
    \centering
    \includegraphics[width=0.95\textwidth]{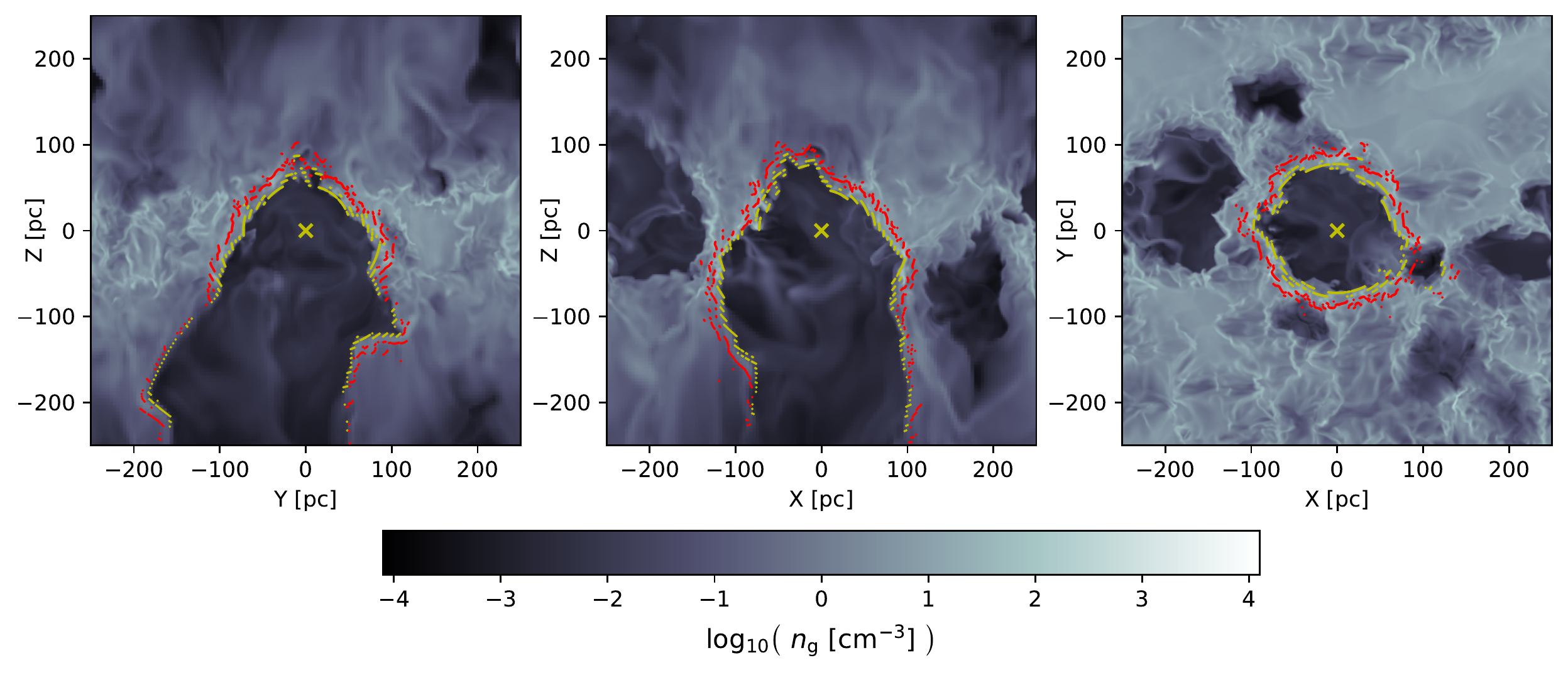}
    
    \caption{Gas number density at the simulation midplanes.
    The yellow and red line segments mark the position of the inner and outer surfaces of the Local Bubble candidate, respectively. The yellow cross represents the observer's position.}
    \label{fig:gasDensFull+edges-lowRes}
\end{figure*}

\begin{figure}

    \centering
    \includegraphics[width=0.48\textwidth]{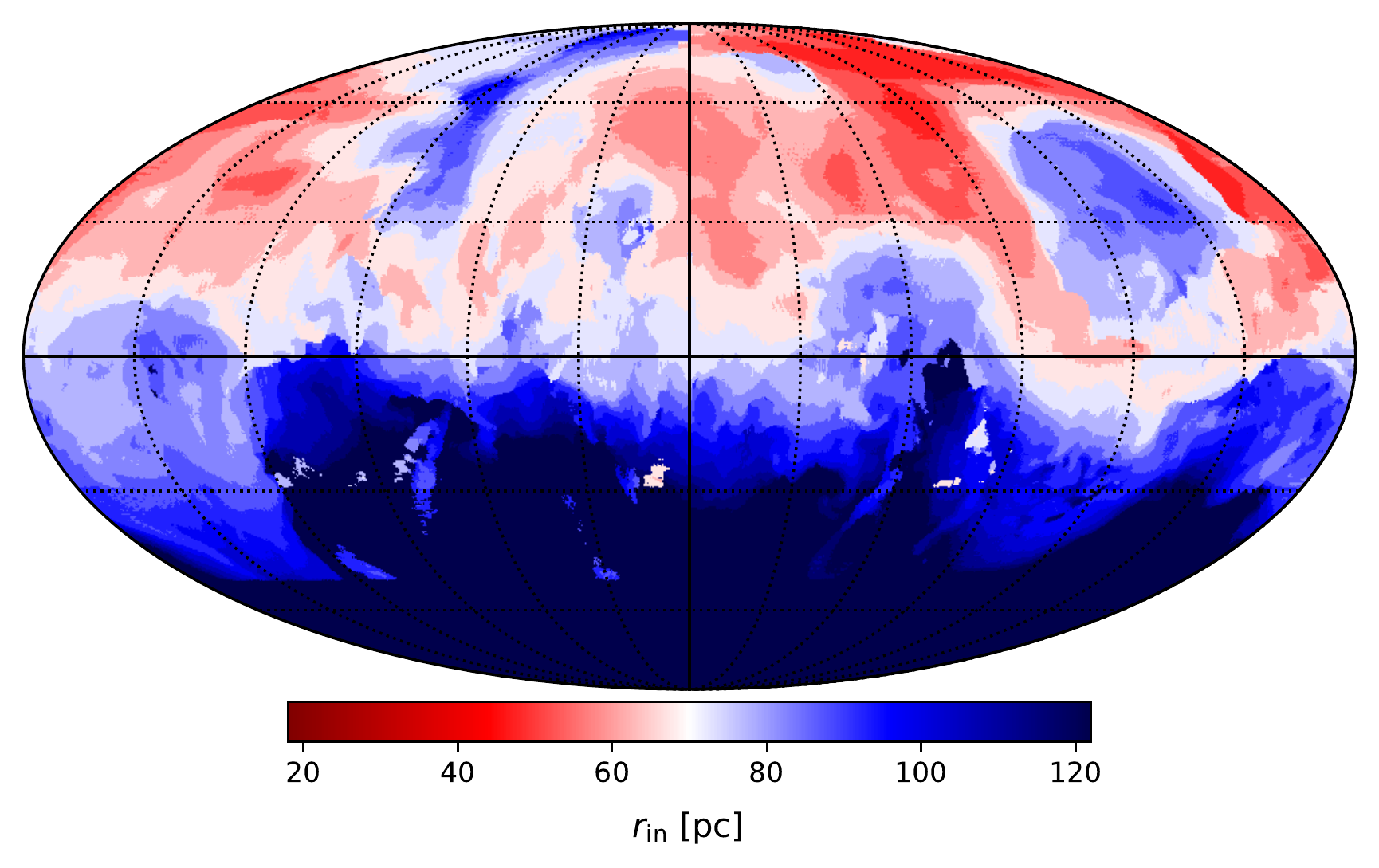}
    
    \caption{Mollweide projection of the distance to the inner surface of the Local Bubble candidate, $r_{\rm in}$, as identified using the method described in Sect.~\ref{subSect:methodWalls}.
    }
    \label{fig:bubbleEdges-lowRes}

\end{figure}

In Fig.~\ref{fig:gasDensFull+edges-lowRes} the gas density is depicted together with the inner and outer surfaces; in Fig.~\ref{fig:bubbleEdges-lowRes} a full-sky map of the distance to the inner surface of the cavity is presented. For the orientation convention of the full-sky map we refer the reader to Sect.~\ref{subSect:conventionMaps}. 

\section{Radiative transfer simulation}\label{sec:synthobs}

All the results presented here are obtained with the same post-processing scheme. 
To compute the dust temperature, the dust grain alignment efficiency and to produce synthetic polarization maps with the relative Stokes' parameters, we make use of the RT code {\tt POLARIS}. We refer the reader to Appendix~\ref{Appendix:POLARIS} and to \cite{Reissl2016} for a more in-depth view of the {\tt POLARIS} RT scheme.

\subsection{Post-processing setup}\label{subSect:post-processing}
As a first step, we calculate with {\tt POLARIS} the dust temperature and the grain alignment efficiency, having given as input the gas density, the gas temperature, the magnetic fields, the dust model, and the sources of emitting photons.
Our dust component is based on a canonical model of the ISM, with a constant dust-to-gas ratio $m_{\rm{dust}}/m_{\rm{gas}}=0.01$ in each grid cell and it consists of a mixture of astronomical silicate ($37.5 \%$) and graphite ($62.5 \%$) materials.
To trace dust polarization, we consider non-spherical dust grains with a size range from $a_{\rm{min}} = 5\,$nm to $a_{\rm{max}}= 200\,$nm and a  size distribution following a power law of $N_{\mathrm{d}}(a) \propto a^{-3.5}$, where $a$ is the effective radius of a non-spherical dust grain equivalent to the volume of a sphere of the same mass \citep{Dohnanyi1969, Mathis1977}. 

As a source of emitting photons, we assume a Mathis' interstellar radiation field \citep[ISRF;][]{Mathis1983, Camps2015} which takes into account the typical stellar emission and the dust thermal radiation in the Milky Way as well as the contribution of the CMB. For the ISRF we assume a strength factor $G_0 = 1.6 $ with $2 \cdot 10^{8}$ photon packages, which is a good compromise between run-time and noise.

As a second step, we perform a ray-tracing simulation at a frequency of 353\,GHz  for an observer placed at the very centre of the cavity. 
We use the same setup adopted to find the bubble walls: 70 concentric spheres spaced by 5\,pc apart are used to sample the celestial sphere according to the  {\tt HEALPix} pixelisation \citep[][$N_{\rm side}=256$]{Gorski2005}. 
{\tt POLARIS} returns, for each LOS, the column density, the optical depth, and the relative Stokes' vector $\textbf{S}=(I,Q,U,V)^{T}$, where $I$ is the total intensity, $Q$ and $U$ represent the linear polarization, $V$ the circular polarization. For each LOS we can therefore determine the linear polarized intensity $P_{\mathrm{l}}$, the degree of linear polarization $p_{\mathrm{l}}$ and the polarization angle $\psi$,

\begin{equation}
    \label{eqn:StokesParam}
        P_{\mathrm{l}} = \sqrt{Q^2+U^2}, \quad  p_{\mathrm{l}} = \frac{P}{I}, \quad \psi = \frac{1}{2}\tan^{-1} \Big(\frac{U}{Q}\Big).
\end{equation}
We emphasize that at 353\,GHz the dust polarization is perpendicular to the magnetic field and $\psi$ needs to be rotated by $90^{\circ}$ to infer the magnetic field orientation. 
\subsection{Line of sight analysis technique}\label{subSect:LOS-analysis}

The transport of radiation is a 3D problem but, unfortunately, in any astronomical observation of radiation emitted by the dust one of the spatial dimensions is lost. The advantage of synthetic observations made with an RT post-processing technique is that the 3D information remains accessible along each LOS and, as a consequence, the origin of the polarized signal can be explored in hindsight.
The procedure is similar to that introduced in \cite{Reissl2020-spiralArm}. 
To track the polarization signal, we  defined a series  of concentric {\tt HEALPix} detectors; in such a way, we can evaluate the first derivative of the signal along each path element $\mathrm{d}r$ of the LOS as
\begin{equation}
    \Delta x_{\mathrm{i}}(r) = \frac{x_{\mathrm{i}}(r+\mathrm{d}r)-x_{\mathrm{i}}(\mathrm{d}r)}{\mathrm{d}r},
\end{equation}
where $x$ can represent e.g. the intensity or the polarized intensity, and the index $i$ stands for a distinct position along a certain LOS, i.e. a pixel of the  {\tt HEALPix} map. Using the just defined quantity and computing its cumulative, we can evaluate the accumulated value $x$ up to a given distance 
\begin{equation}
    x_{\mathrm{i}} (r) = \int_{0}^{r} \Delta x_{\mathrm{i}}(r') \, \mathrm{d}r',
\end{equation}
where $r$ is the distance from the observer.
We study the origin of the polarization signal along all LOSs and the contribution  of the local environment, i.e. the cavity walls, to the properties of the received light.

\subsection{Mollweide and orthographic projections orientation}\label{subSect:conventionMaps}

We present full-sky maps and orthographic views of the data whose orientation is decided and defined as follows.

The orientation of the maps is constrained  by the initial magnetic field orientation of our simulation (parallel to the $x$-axis; see Sect.~\ref{subSec:simul+select-bubble}) and by the one observed in our location in the Milky Way.
We therefore set the centre of the full sky-maps, $(l,b)=(0^{\circ},0^{\circ})$, to coincide  with the LOS that propagates from the observer toward the $-y$ direction in the $x-y$ plane; here $l$ and $b$ are the longitude and the latitude, respectively. The $-y$ direction can be considered as the LOS toward the hypothetical Galactic centre of our simulation. Moving from $(l,b)=(0^{\circ},0^{\circ})$ to the left (positive increase in longitude) is equivalent to moving in a clockwise direction in the $x-y$ plane from the starting LOS; moving from centre to the right is instead equivalent to moving in a counter-clockwise direction.

In the orthographic projections, the left and right panels touch at  $l=-90^{\circ}$; $l=0^{\circ}$ coincide with the  lower point of the circle.

\section{Results and Discussion}\label{sec:discussion}

We expect the walls of the cavity to have an impact on the polarization observations performed inside it, since dust will act  both as a polarizing filter and as a source of polarized light alike. Indeed, compared to the larger scale environment, the bubble edges have a greater dust density, as a natural result of the cavity expansion; moreover, they are crossed by magnetic field lines that are expected to be coherent at least at latitudes smaller than $65^\circ$. This coherence may not be observable at the poles ($|b|>65^\circ$), where the expansion can open the bubble causing the tangling of field lines. We also expect enhanced magnetic field strength in the cavity walls (see second panel of Fig.~\ref{fig:allCuts-lowRes}) due to the squeezing of the field lines into a thin shell. Such a cold, dense, and dusty environment crossed by strong and, in some places, coherent magnetic field lines is fundamental for having dust grain alignment and subsequent polarization \citep[see e.g.][]{Hughes2009, Hoang2009}. 

To study the influence of the bubble over polarization we produce maps of synthetic Stokes' parameters for three different setups: the original simulation (S0), the case in which only the part enclosed by the outer surface of the bubble is considered (S0justLB), and the case where the magnetic field is set only along the $x$ direction (S0uniB) with a strength equal to the mean value of the magnetic field  \textbf{B} of the simulation.  In Table~\ref{tab:simOverview} we list an overview of the RT post-processing simulations. 

\begin{table}
	\centering
	\caption{Synthetic observations overview.}
	\label{tab:simOverview}
	\begin{tabular}{lll} 
		\hline
		Sim. name & Density field & Magnetic field\\
		\hline
		S0       & original  & original\\
		S0justLB & enclosed by the outer edge  & original\\
		S0uniB   & original  & \textbf{B}=(1.8$\,{\rm \mu G}$) $\hat{x}$ \\
		\hline
	\end{tabular}
\end{table}

In this section, we discuss in detail the influence that the presence of the cavity has on the column density ($N_{\rm H}$), on the linear polarized flux ($P_{\mathrm{l}}$), and on the linear polarization fraction ($p_{\mathrm{l}}$) as received by an observer placed inside the bubble. We present the magnetic field lines direction as inferred by the dust polarization angles at 353\,GHz and we compare them qualitatively with those observed for the Milky Way by the {\it Planck} mission. We end the discussion by briefly comparing an open cap with a closed one and pointing out some limits of our work.

\subsection{Cavity and observed polarized emission}

\begin{figure*}
    \centering
     \includegraphics[width=0.95\textwidth]{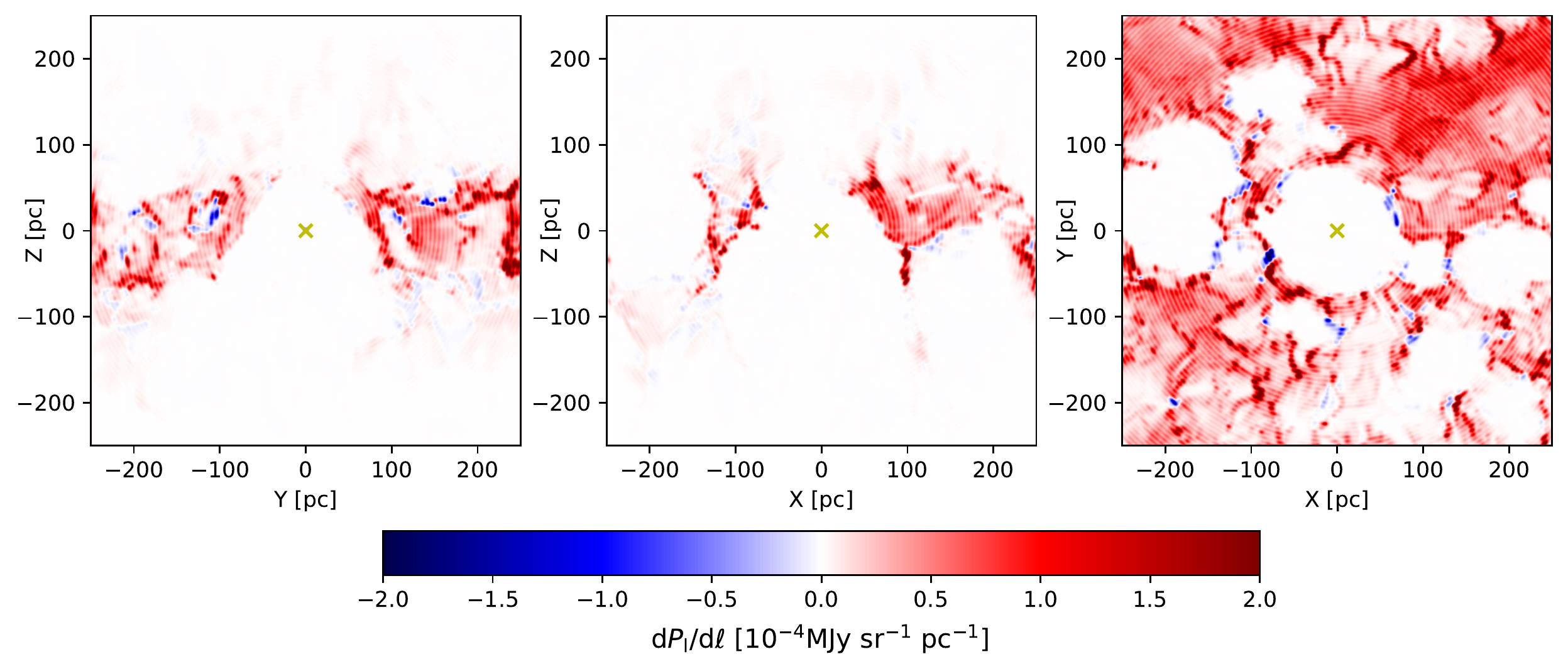}
    \caption{Same as Fig.~\ref{fig:allCuts-lowRes}, but for the first derivative of the linearly polarized flux at 353\,GHz along the LOS, $\mathrm{d}P_{353}/\mathrm{d}r$\,$=$\, $(P_{353}(r+\mathrm{d}r)-P_{353}(r))/\mathrm{d}r$, calculated in the synthetic observation produced using {\tt POLARIS}.
    The red (blue) colours correspond to increases (decreases) in $P_{353}$ along each LOS toward an observer at the position at the centre marked by a yellow cross, respectively. The concentric ring regular pattern visible in the images is due to the finite number of detectors used to sample the rays of light.}
    \label{fig:polFluxChange-lowRes}
\end{figure*}

In Fig.~\ref{fig:polFluxChange-lowRes} we present the first derivative of the linearly polarized flux, at 353\,GHz, along all the LOSs that lie within the midplanes as they approach the observer from outside of the simulation domain. From the comparison of these cuts with the corresponding gas density cuts (dust to gas ratio is $1\,\%$), in the top panel of Fig.~\ref{fig:allCuts-lowRes}, it is clearly visible that the linearly polarized dust emission is correlated to the dust density. Inside the simulated cavities, where the material has been swept away by the energy of SN explosions, there is negligible dust emission; this emission becomes instead  important in the densest parts of the simulation, specifically at the cavity walls where magnetic fields are stronger and coherent.

\begin{figure*}
    \centering
    \includegraphics[width=.8\textwidth]{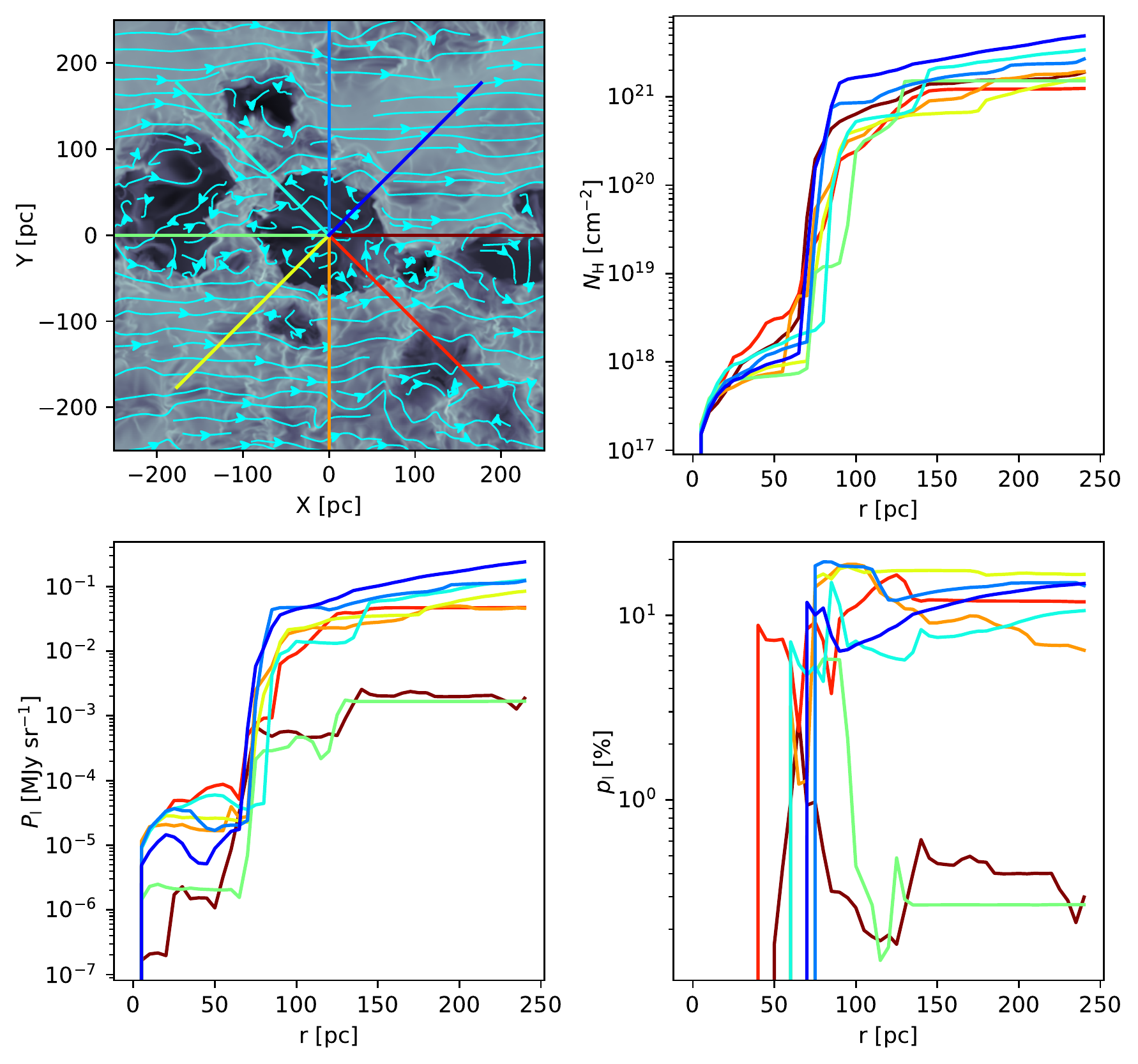}
    \caption{Column density ($N_{\rm H}$), linearly polarized flux ($P_{\rm l}$), and linear polarization fraction ($p_{\rm l}$) over distance, as seen by an observer placed at the centre of the cavity, along eight LOSs in the $x-y$ plane. The eight LOSs chosen are presented in the the upper left panel and are the same of Fig.~\ref{fig:differential_NH}.}
    \label{fig:polRays_XYplane-lowRes}
\end{figure*}

We use the {\tt POLARIS} ray-tracing simulation to sample the Stokes' parameters and the column density values while the rays are approaching the observer at the centre of the bubble, as described in Sect.~\ref{subSect:LOS-analysis}.
In Fig.~\ref{fig:polRays_XYplane-lowRes} we report the value of $N_{\rm H}$, $P_{\mathrm{l}}$, and $p_{\mathrm{l}}$ with respect to the distance from the centre for eight selected LOSs chosen in the $x-y$ plane and spaced by an angle of 45$^{\circ}$. Starting from $r=0 \, {\rm pc}$ and moving away from the centre, these three quantities undergo an abrupt change in their values as the inner edge of the cavity is crossed. Moving further through the cavity wall more material is taken into account and this results in a constant increase of  $N_{\rm H}$; instead, the values of $P_{\mathrm{l}}$ and $p_{\mathrm{l}}$ change more randomly since they depend on the coherence of the magnetic field lines orientation along the LOS as well as the local dust grain alignment efficiency \citep[see e.g.][]{Lazarian2007,Reissl2020}.

\begin{figure*}
    \centering
    \includegraphics[width=0.9\textwidth]{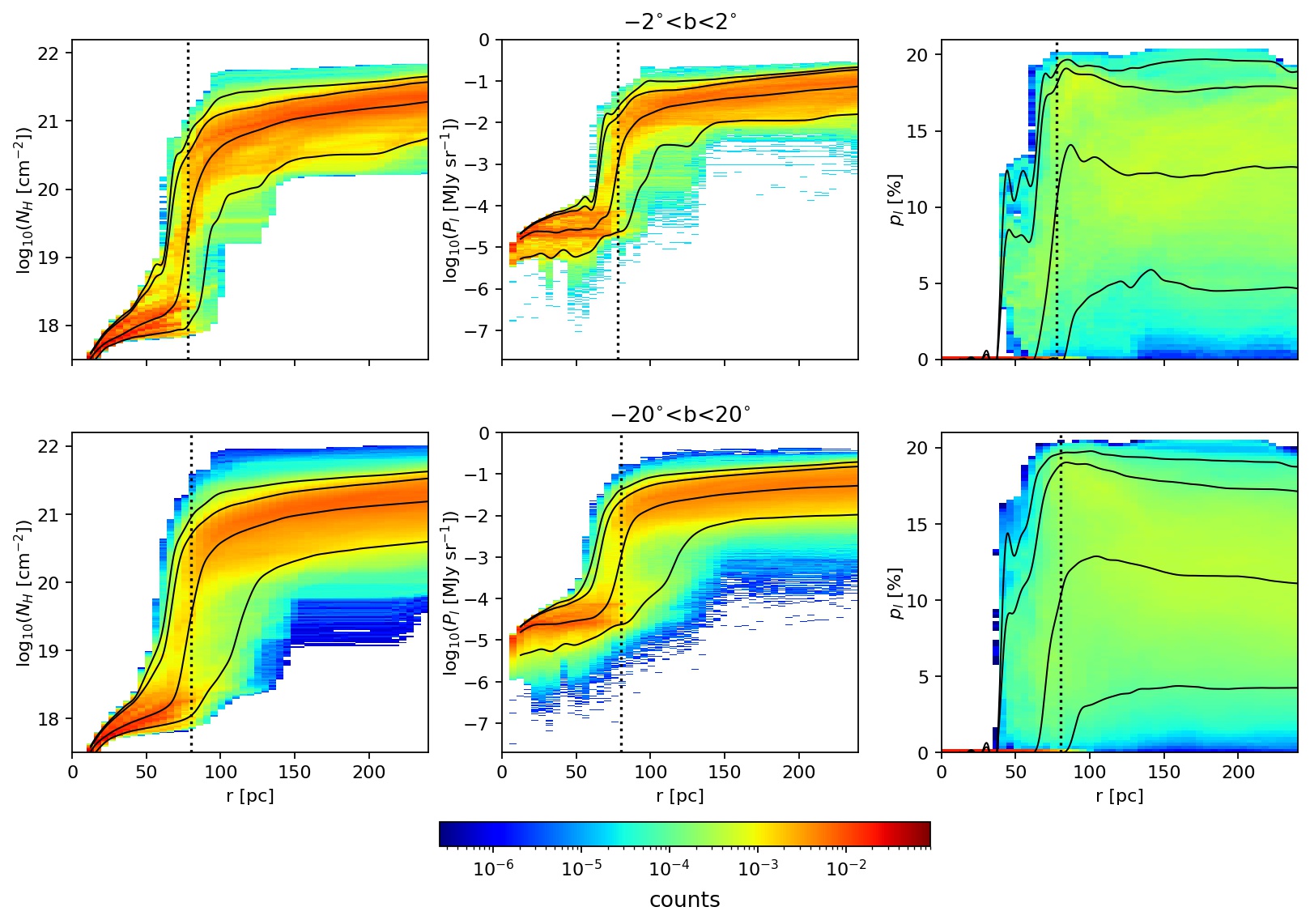}
    \caption{Two-dimensional normalized histograms for the column density, the linearly polarized flux, and the linear polarization fraction for all the LOSs that lie within the Galactic latitude range indicated in each row. 
    The black solid lines show the 5th, the 95th and 99th percentiles of the data and the median. 
    The black dotted vertical lines indicate the average distance to the inner surface of the Local Bubble candidate.}
    \label{fig:2Dhist_XYplane-lowRes}
\end{figure*}

In Fig.~\ref{fig:2Dhist_XYplane-lowRes} we report the normalized two-dimensional histograms for all the LOSs that lie within a given latitude interval ($b \in$ $\rm{(}-2^\circ,\, 2^\circ \rm{)}$; $\rm{(}-20^\circ,\, 20^\circ \rm{)}$). The 5th, 95th, and 99th percentiles, together with the median, are plotted. The mean distance to the inner surface of the cavity is reported as a vertical line. 
In these plots we find a confirmation of the influence of the edges of the bubble on  $N_{\rm H}$, $P_{\mathrm{l}}$, and $p_{\mathrm{l}}$, since these quantities have a consistent variation in their values in correspondence of the cavity borders. 

We isolate the contribution from the bubble edges to $N_{\rm H}$, $I$, $P_{\mathrm{l}}$, and $p_{\mathrm{l}}$ by considering the S0justLB setup.
We report the resulting values for S0 and S0justLB in Table~\ref{tab:comparison}.
The comparison between the obtained values indicates that the contribution of the edges of the Local Bubble is not negligible with respect to that of the whole simulation.

We know from real observations that the Bubble boundary can be identified as the distance from the observer where the linear polarization fraction in the $V$ band reaches a value of $\approx 0.1$\% \citep{Gontcharov2019}. To check if our simulations are in agreement with observations, we use the S0justLB setup and the relation presented in \cite{Planck2015-XXI},

\begin{equation}
    R_{\mathrm{P/p}} = \frac{P_{\rm{S}}}{p_{\rm{V}}} = [5.4 \pm 0.3] \, \rm{MJy \ sr^{-1}},
\end{equation}
where $P_{\rm{S}}$ is the linear polarized flux at $353 \, \rm{GHz}$ and $p_{\rm{V}}$ is the polarization fraction in the visible band.
This relation  correlates the polarization properties in the sub-millimetre band with the ones in the visible. We obtain a median threshold value for $p_V$ of $\approx 0.12$\%, which is comparable with the work of \cite{Gontcharov2019}.

\begin{table*}
	\centering
	\caption{Comparison of the column density, intensity, linearly polarized flux, and polarization fraction between S0 and S0justLB. The quantities are computed using all the LOSs and only the LOSs at low and intermediate latitudes ($15^{\circ} \le |b| \le 65^{\circ}$).}
	\label{tab:comparison}
	\begin{tabular}{llcccc} 
		\hline
		Quantity & Sim. & \multicolumn{2}{|c|}{Mean} & \multicolumn{2}{|c|}{Median} \\
		& & all & $15^{\circ} \le |b| \le 65^{\circ}$ & all & $15^{\circ} \le |b| \le 65^{\circ}$ \\
		\hline
		
		${\rm{log}_{10}}(N_{\rm{H}})$ \,[$\rm{cm}^{-2}$] & S0 & 20.97 & 20.76 & 20.72 & 20.60 \\ 
		 & S0justLB & 20.34 & 20.21 & 20.12 & 19.97  \\  
		\hline
		$I$\,[MJy\,sr$^{-1}$] & S0 & 0.324 & 0.210 & 0.198 & 0.151 \\  
		 & S0justLB &  0.082 & 0.063 & 0.05 & 0.036 \\  
		\hline
		$P_\mathrm{l}$\,[MJy\,sr$^{-1}$] & S0 & 0.032 & 0.016 & 0.013 & 0.009 \\  
		 & S0justLB & 0.009 & 0.006 & 0.005 & 0.003 \\  
		\hline
		$p_\mathrm{l}$\,[\%] & S0 & 8.093 & 7.214 & 7.794 & 6.913 \\  
		 & S0justLB & 9.991 & 9.621 & 10.081 & 9.517 \\ 
		\hline
	\end{tabular}
\end{table*}

\subsection{Cavity and polarization angle}

\begin{figure*}
    \centering
    \begin{subfigure}[b]{.45\textwidth}
        \centering
        \includegraphics[width=1.0\textwidth]{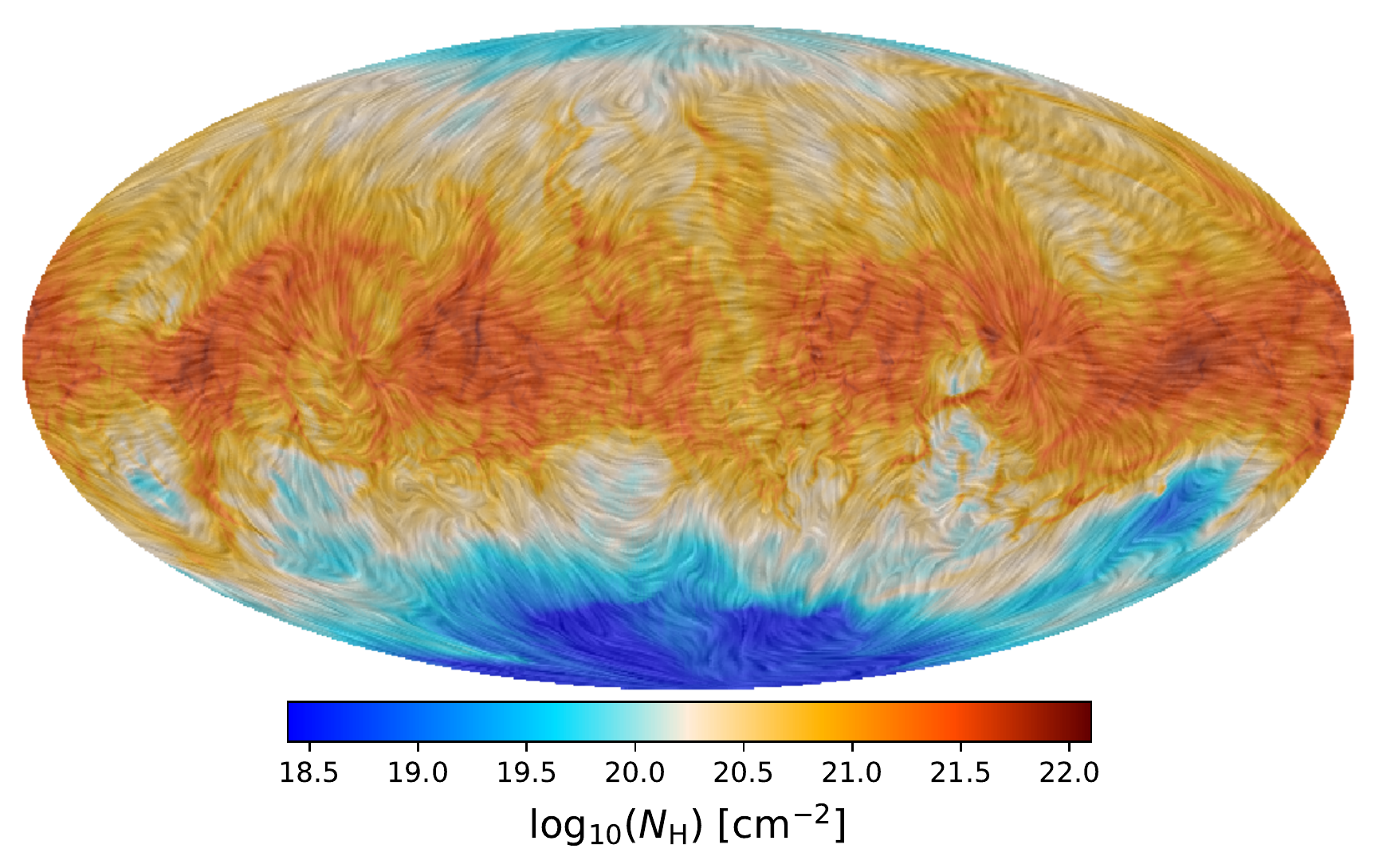}
    \end{subfigure}
    \hfill
    \begin{subfigure}[b]{.45\textwidth}
        \centering
        \includegraphics[width=1.0\textwidth]{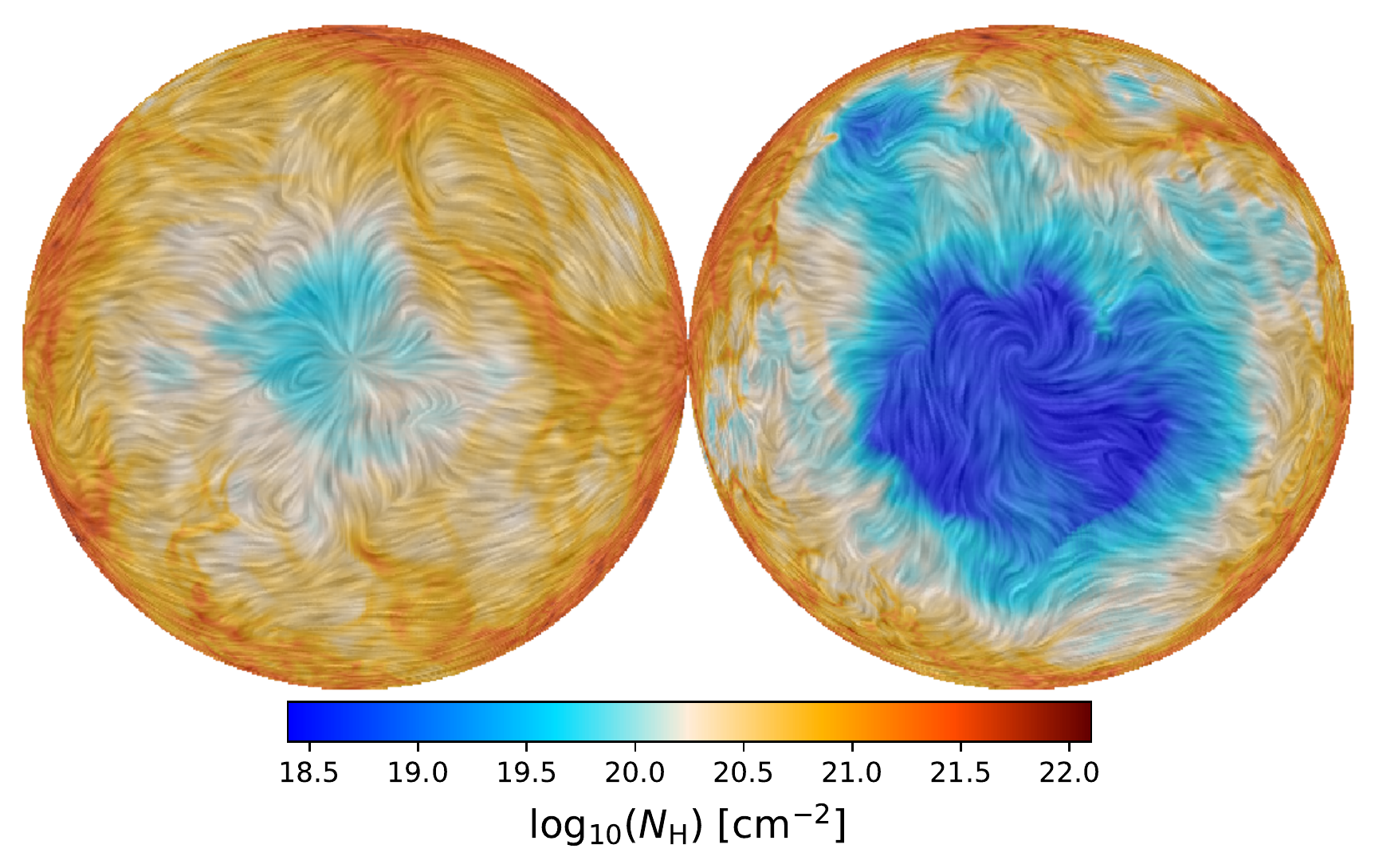}
    \end{subfigure}
    \begin{subfigure}[b]{.45\textwidth}
        \centering
        \includegraphics[width=1.0\textwidth]{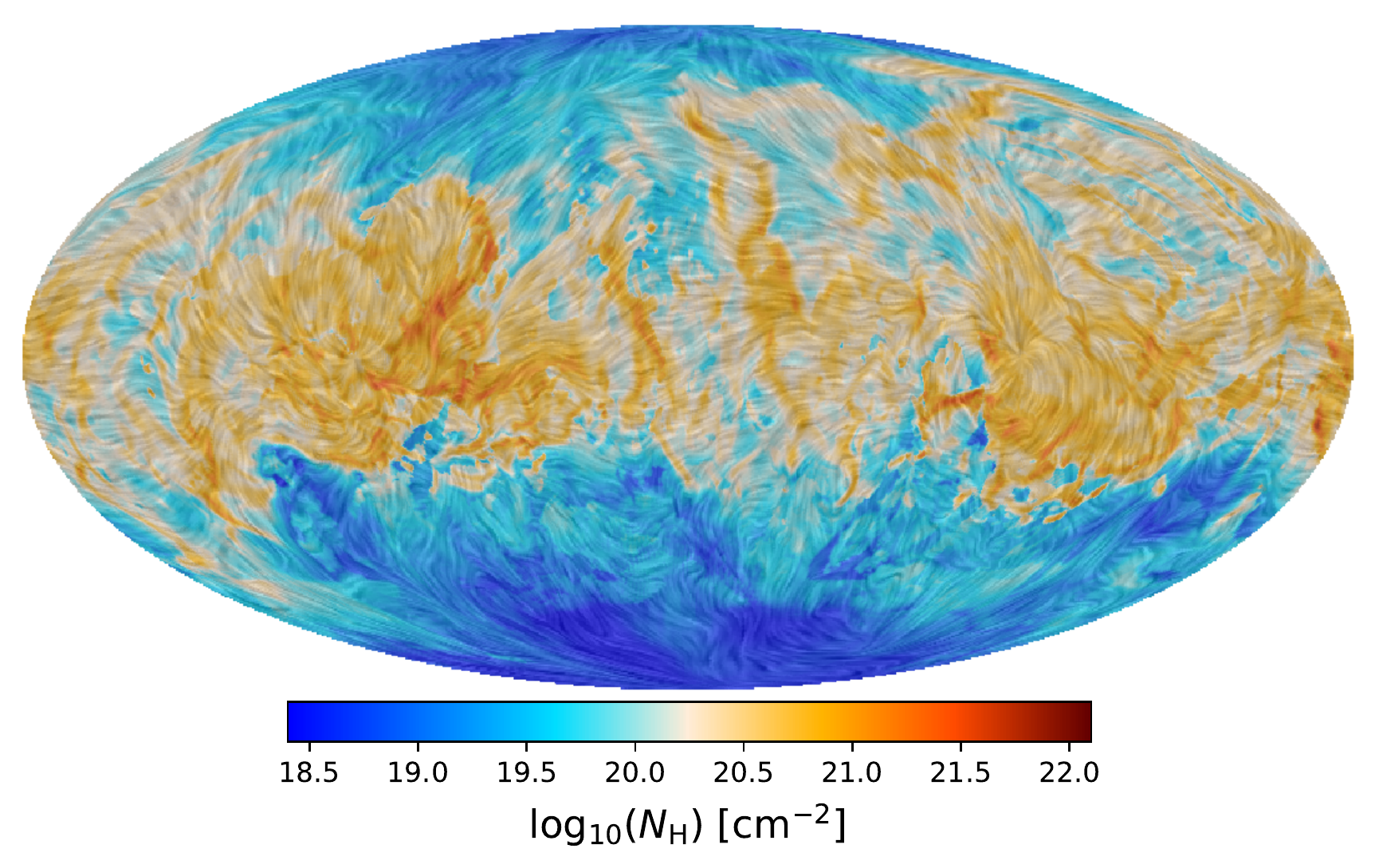}
    \end{subfigure}
    \hfill
    \begin{subfigure}[b]{.45\textwidth}
        \centering
        \includegraphics[width=1.0\textwidth]{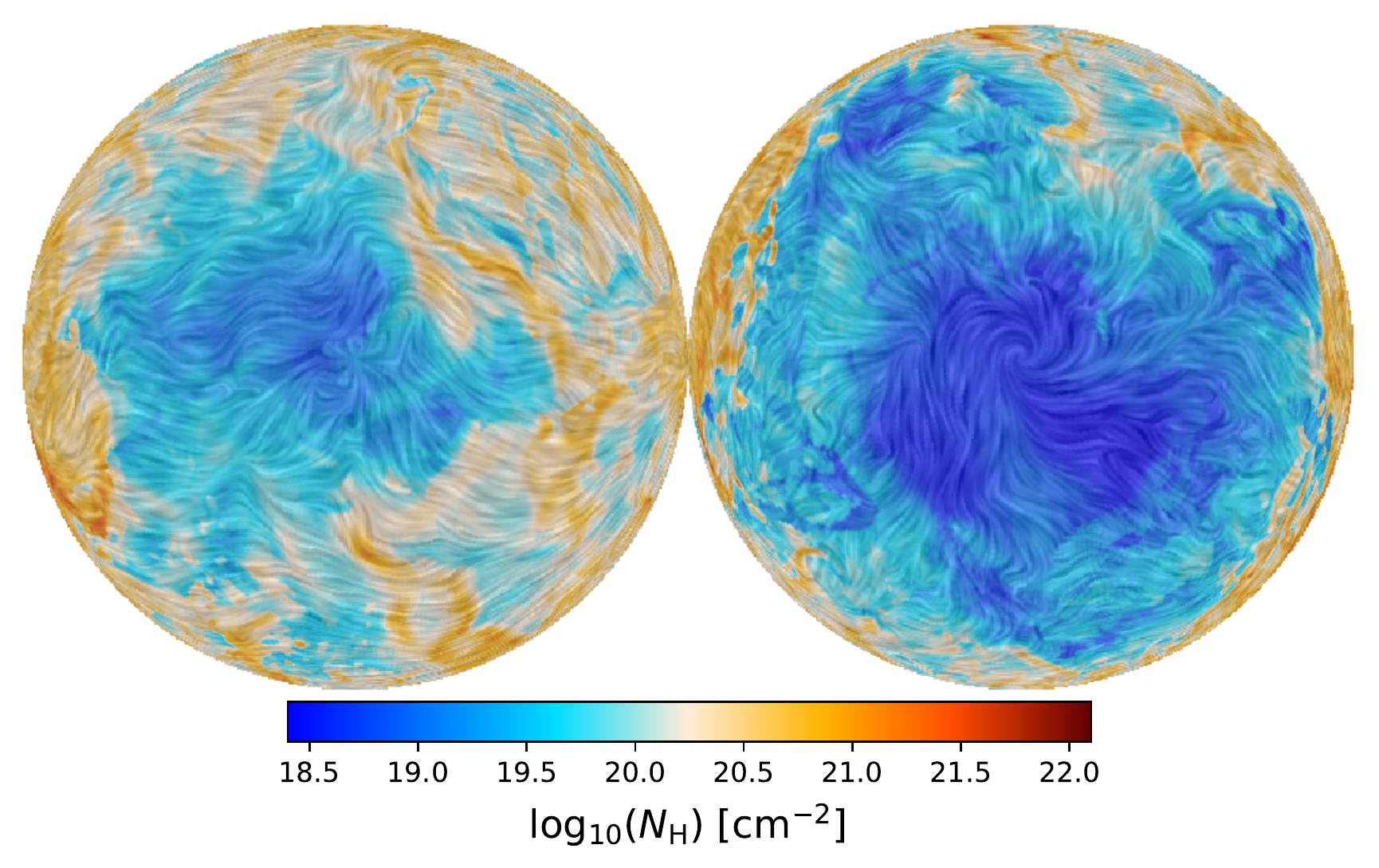}
    \end{subfigure}
     \begin{subfigure}[b]{.45\textwidth}
        \centering
        \includegraphics[width=1.0\textwidth]{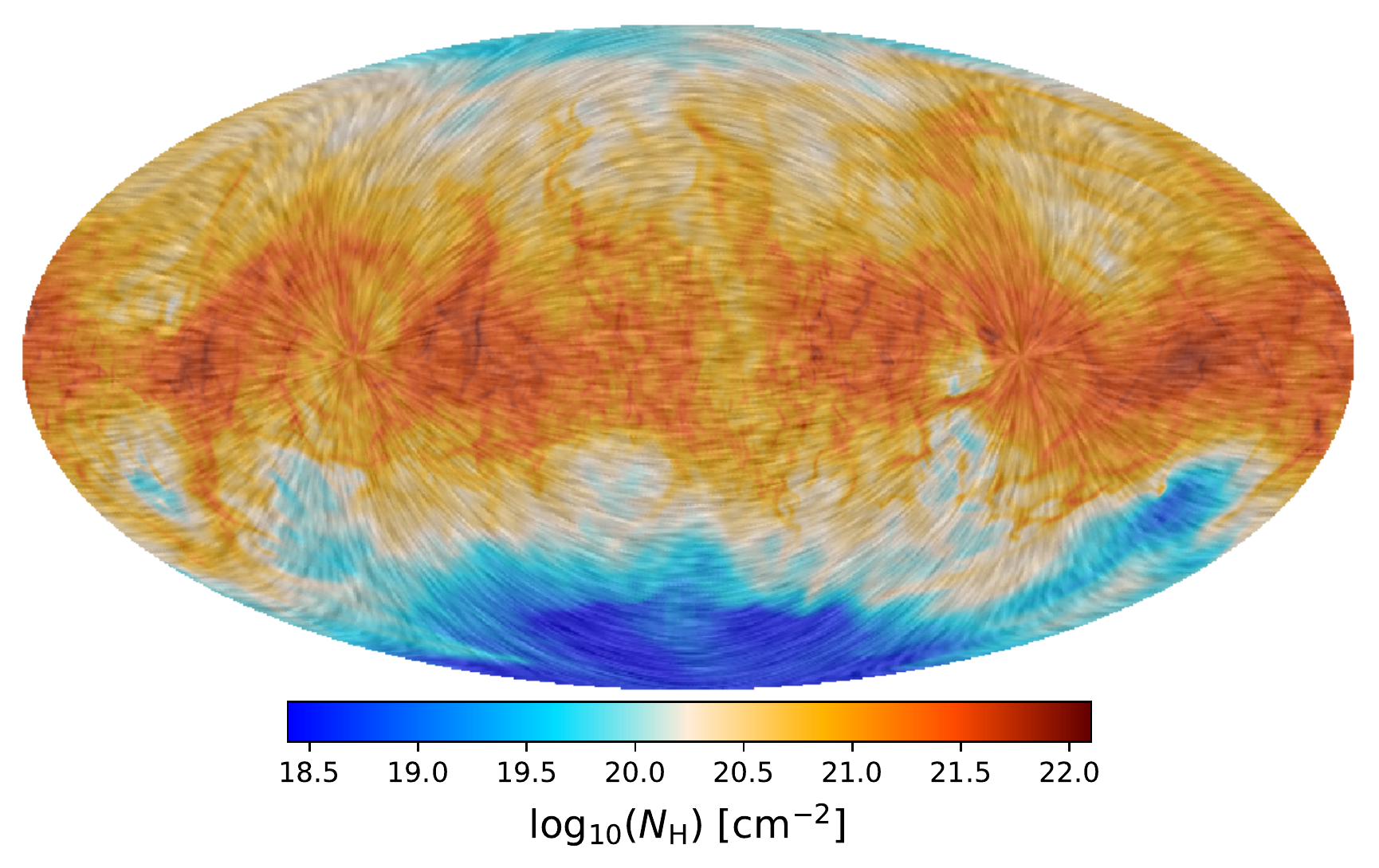}
    \end{subfigure}
    \hfill
    \begin{subfigure}[b]{.45\textwidth}
        \centering
        \includegraphics[width=1.0\textwidth]{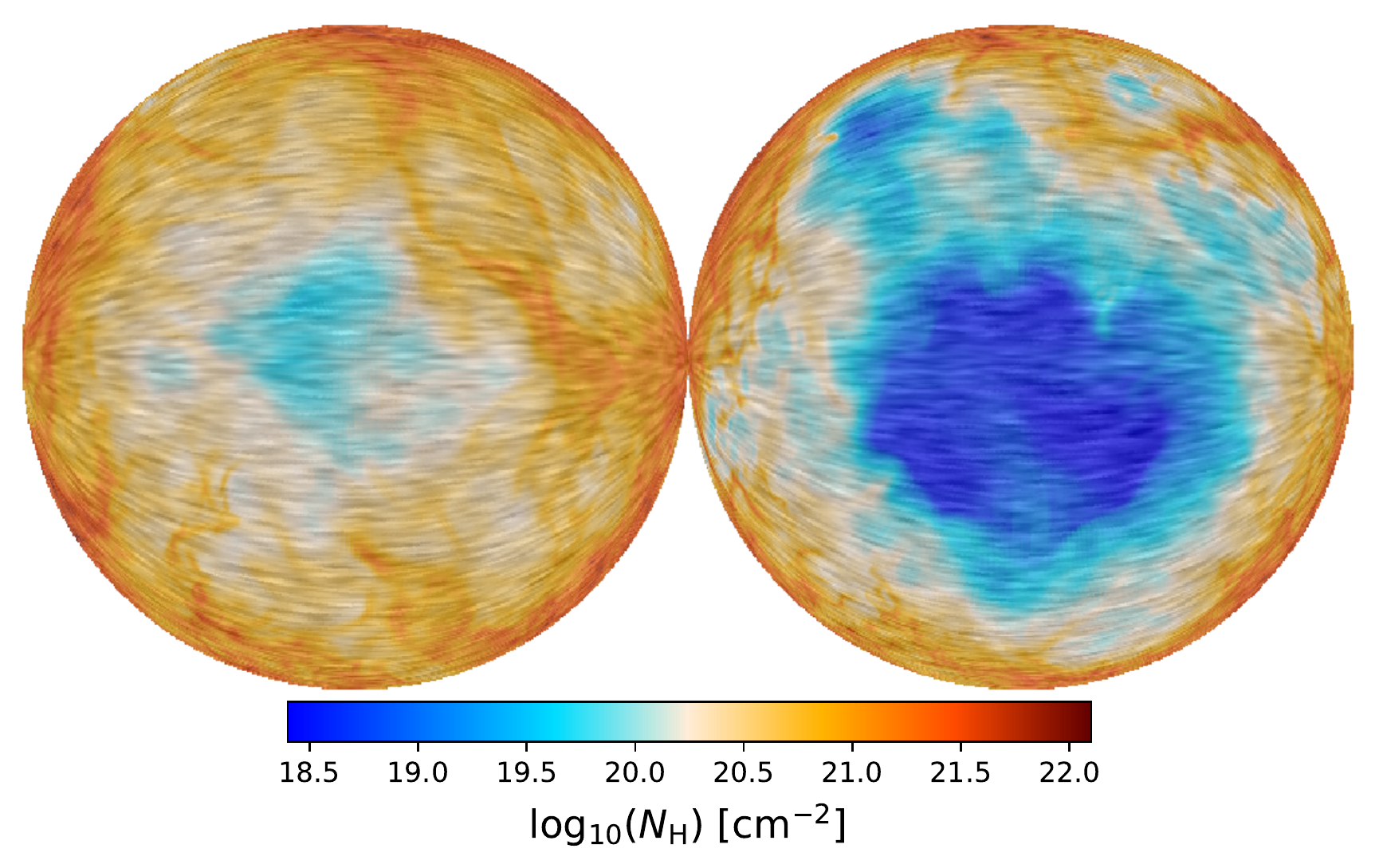}
    \end{subfigure}
    \caption{
    Mollweide (\textit{left}) and orthographic projections centered at the Galactic poles (\textit{right}) of the column density, $N_{\rm H}$, and the plane-of-the-sky magnetic field derived from the synthetic 353\,GHz polarization, shown by the colours and the drapery pattern, respectively. The drapery pattern is obtained using the line integral convolution \citep[LIC;][]{cabral1993} of the rotated dust polarization vectors.
    From top to bottom, the panels correspond to the synthetic observations S0, S0justLB, and S0uniB, as introduced in Table~\ref{tab:simOverview}.}
    \label{fig:maps-NH+LIC-lowRes}
\end{figure*}

We use the Stokes parameters from the RT simulations to compute the polarization angles from the dust emission, which we rotate by 90$^{\circ}$ to obtain the plane-of-the-sky magnetic field orientation.
Figure~\ref{fig:maps-NH+LIC-lowRes} presents Mollweide and orthographic projections centered at the Galactic poles for the column density, $\log_{10}(N_{\rm{H}})$, and the inferred magnetic field orientation for the three setups described in Table~\ref{tab:simOverview}.
We consider the results into three distinct regions of the sky: the disk midplane ($|b|$\,$<$\,$15^{\circ}$), low- to intermediate latitudes ($15^{\circ}$\, $\le$\,$|b|$\,$\le$\,$65^{\circ}$), and the Galactic polar regions ($|b|$\,$>$\,$ 65^{\circ}$).

The Mollweide projection of S0, presented in the top panel of Fig.~\ref{fig:maps-NH+LIC-lowRes}, shows that the magnetic field lines toward the disk midplane and at low- and intermediate Galactic latitudes are relatively homogeneous and parallel to the Galactic plane.
This homogeneity and orientation are also observed in S0justLB, shown in the middle panel of Fig.~\ref{fig:maps-NH+LIC-lowRes}, where only the volume enclosed by the Local Bubble candidate is considered.
In both the setups, we find the convergence of the magnetic field lines toward two points in the Galactic plane, coinciding with the directions of $(l,b) \approx (\pm90^\circ,0^\circ)$.

We interpret the similarity between S0 and S0justLB at low and intermediate Galactic latitudes to indicate that the cavity in the simulation inherits the initial mean field orientation from its environment.
This is in agreement with theoretical models that assume flux-freezing, i.e. perfect coupling between the matter and the field, to model the magnetic field morphology in the surface of an expanding bubble \citep[see e.g.][]{FerriereEtAl1991,alves2018,PelgrimsEtAl2020}. 
This assumption, added to the general geometry of the expanding shell, results in a relatively homogeneous field parallel to the wall of the cavity and converging toward the poles set by the initial field orientation.
However, at high Galactic latitudes, there is a clear discrepancy between our synthetic observations and the homogeneous field lines expected from theoretical models, as we discuss later.

\begin{figure*}
    \centering
    
    \begin{subfigure}[b]{.45\textwidth}
        \centering
        \includegraphics[width=1.0\textwidth]{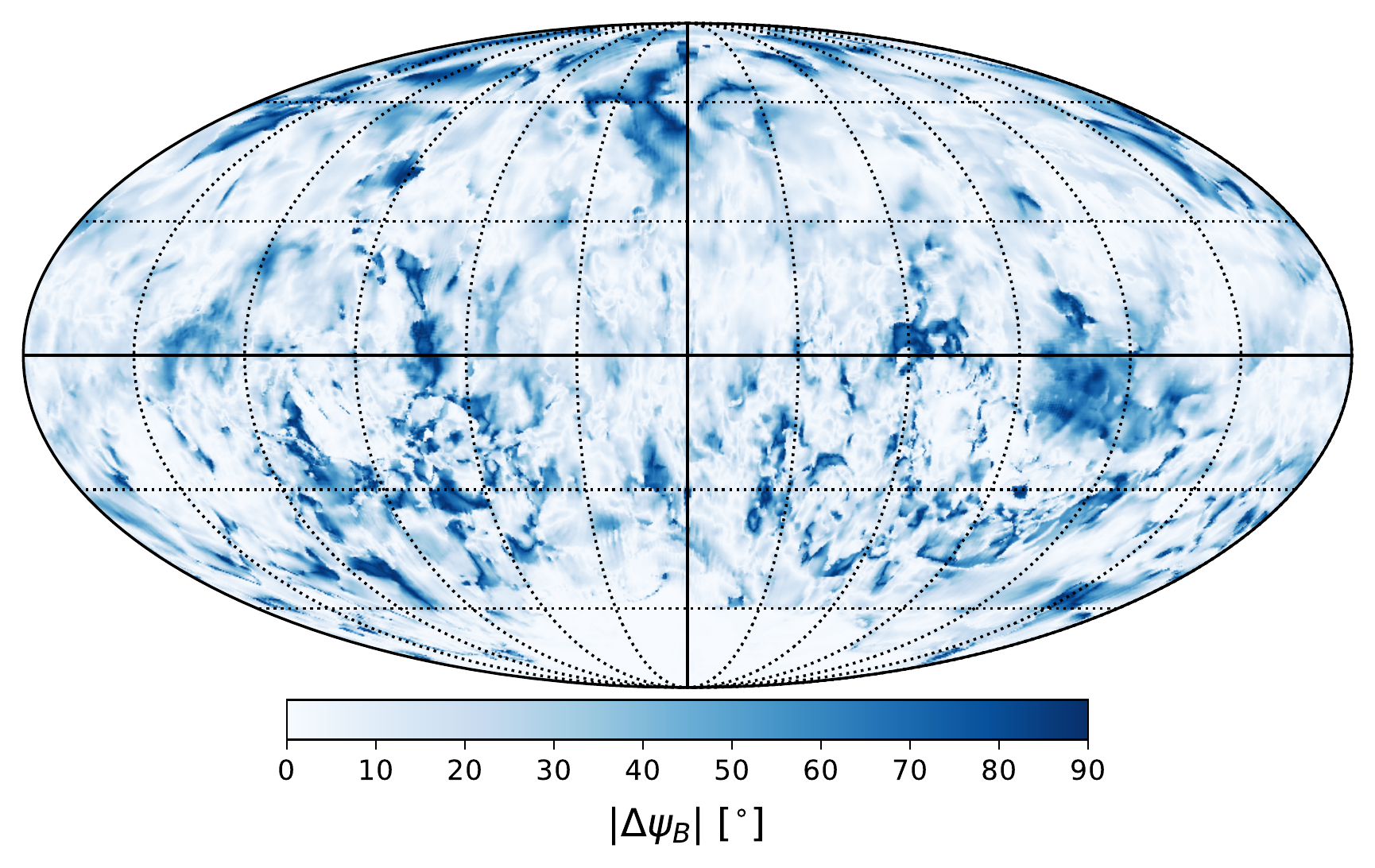}
    \end{subfigure}
    \hfill
    \begin{subfigure}[b]{.45\textwidth}
        \centering
        \includegraphics[width=1.0\textwidth]{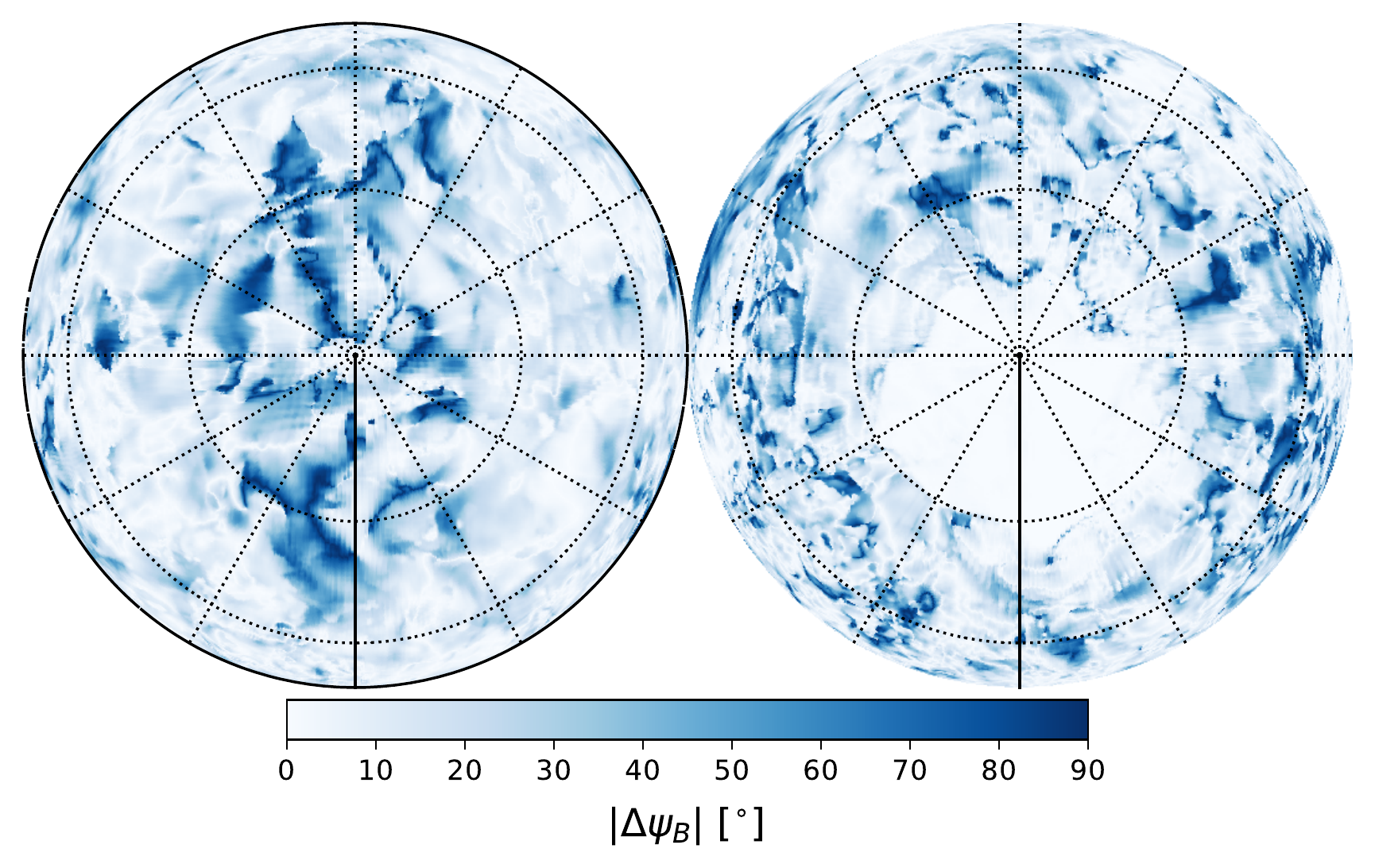}
    \end{subfigure}
    \begin{subfigure}[b]{.45\textwidth}
        \centering
        \includegraphics[width=1.0\textwidth]{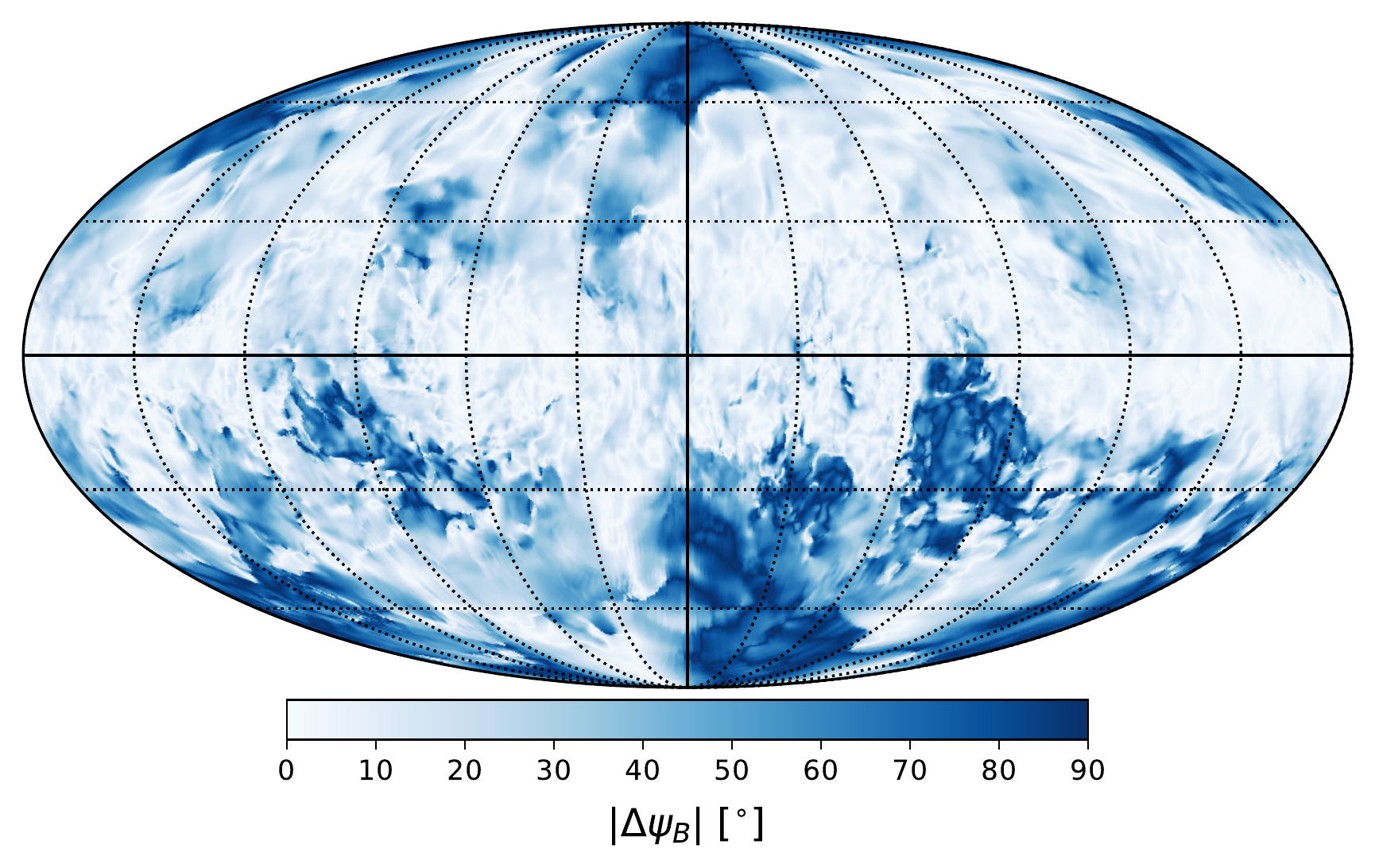}
    \end{subfigure}
    \hfill
    \begin{subfigure}[b]{.45\textwidth}
        \centering
        \includegraphics[width=1.0\textwidth]{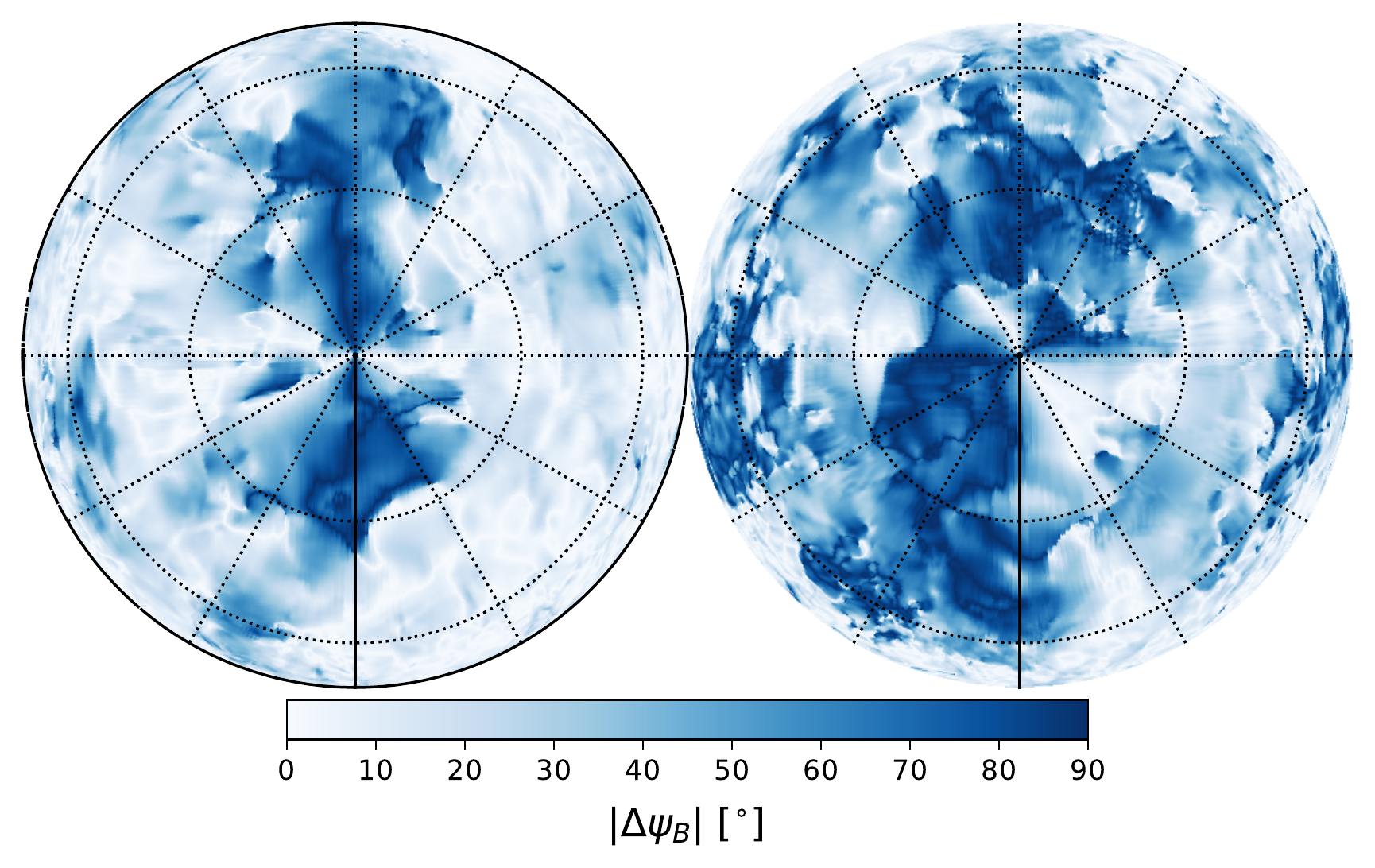}
    \end{subfigure}
     \begin{subfigure}[b]{.45\textwidth}
        \centering
        \includegraphics[width=1.0\textwidth]{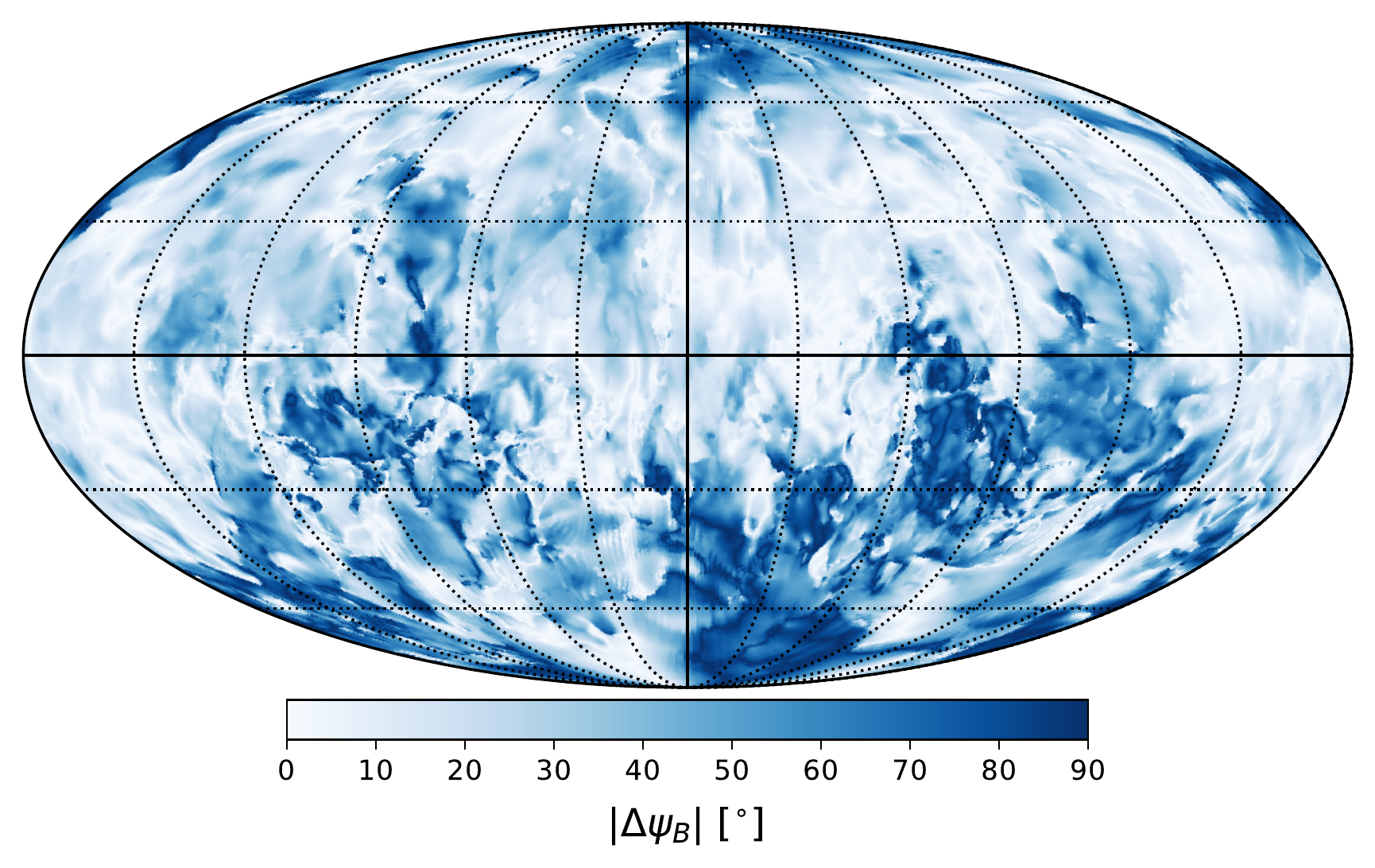}
    \end{subfigure}
    \hfill
    \begin{subfigure}[b]{.45\textwidth}
        \centering
        \includegraphics[width=1.0\textwidth]{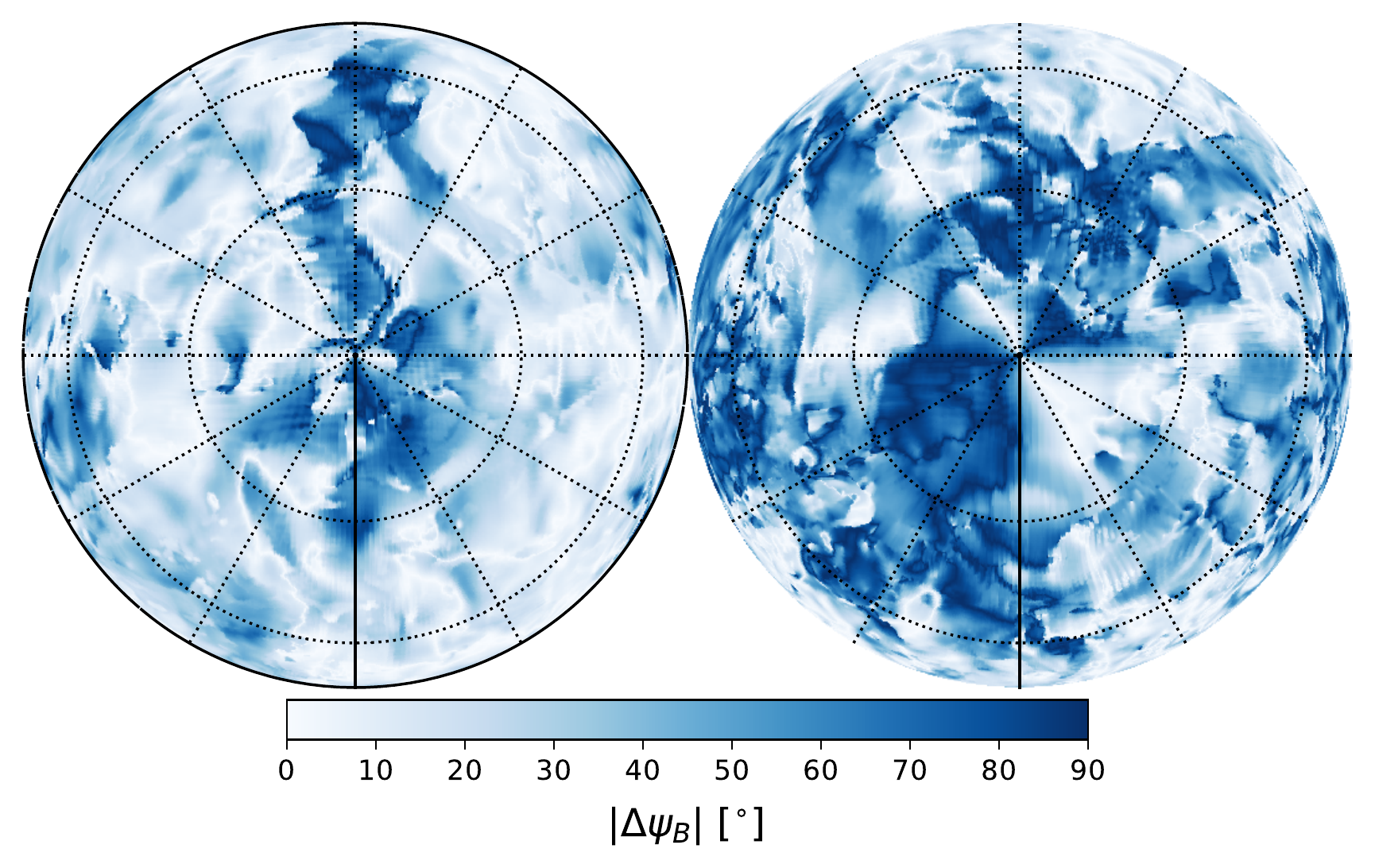}
    \end{subfigure}

    \caption{Mollweide (\textit{left}) and orthographic projections (\textit{right}) of the dust polarization angle difference corresponding to the magnetic field orientation presented in Fig.~\ref{fig:maps-NH+LIC-lowRes}. \textit{Top panel}: difference between the simulations S0 and S0justLB. 
    \textit{Middle panel}: difference between S0 and S0uniB. 
    \textit{Bottom panel}: difference between S0justLB and S0uniB.}
    \label{fig:polAngleDifference-lowRes}
\end{figure*}

We present the differences in dust polarization angles between the three RT setups in Fig.~\ref{fig:polAngleDifference-lowRes}, which illustrate the variations of the polarization pattern among these configurations.
The angle differences between S0 and S0justLB highlight the deviation in the projected field orientation between the Local Bubble candidate and the background.
If the contributions from the cavity to the observed polarization orientation is negligible, we would see large angle differences.
If the candidate Local Bubble is the dominant signal determining the observed polarization orientation, we would see negligible angle differences.
We find that the angle differences are close to 0$^{\circ}$ for most of the sky, but there are a considerable number of LOSs for which the difference is up to 90$^{\circ}$.
This suggests that while the dust and magnetic field in the shell of the Local Bubble candidate plays a fundamental role in creating the observed polarization patterns, it is not solely responsible for the signal across the whole sky.

\begin{figure*}
    \centering

     \begin{subfigure}[b]{.45\textwidth}
        \centering
        \includegraphics[width=1.0\textwidth]{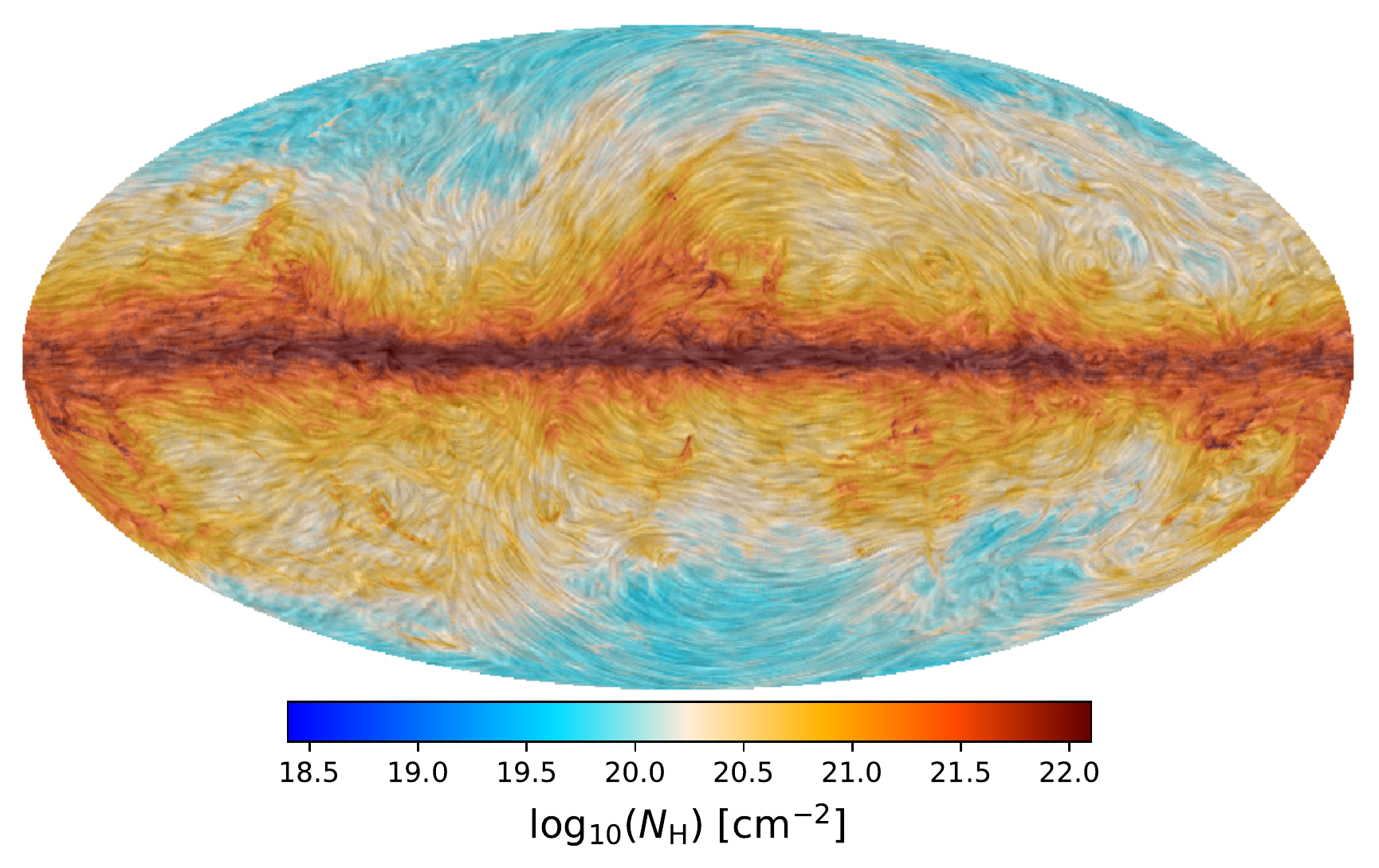}
    \end{subfigure}
    \hfill
    \begin{subfigure}[b]{.45\textwidth}
        \centering
        \includegraphics[width=1.0\textwidth]{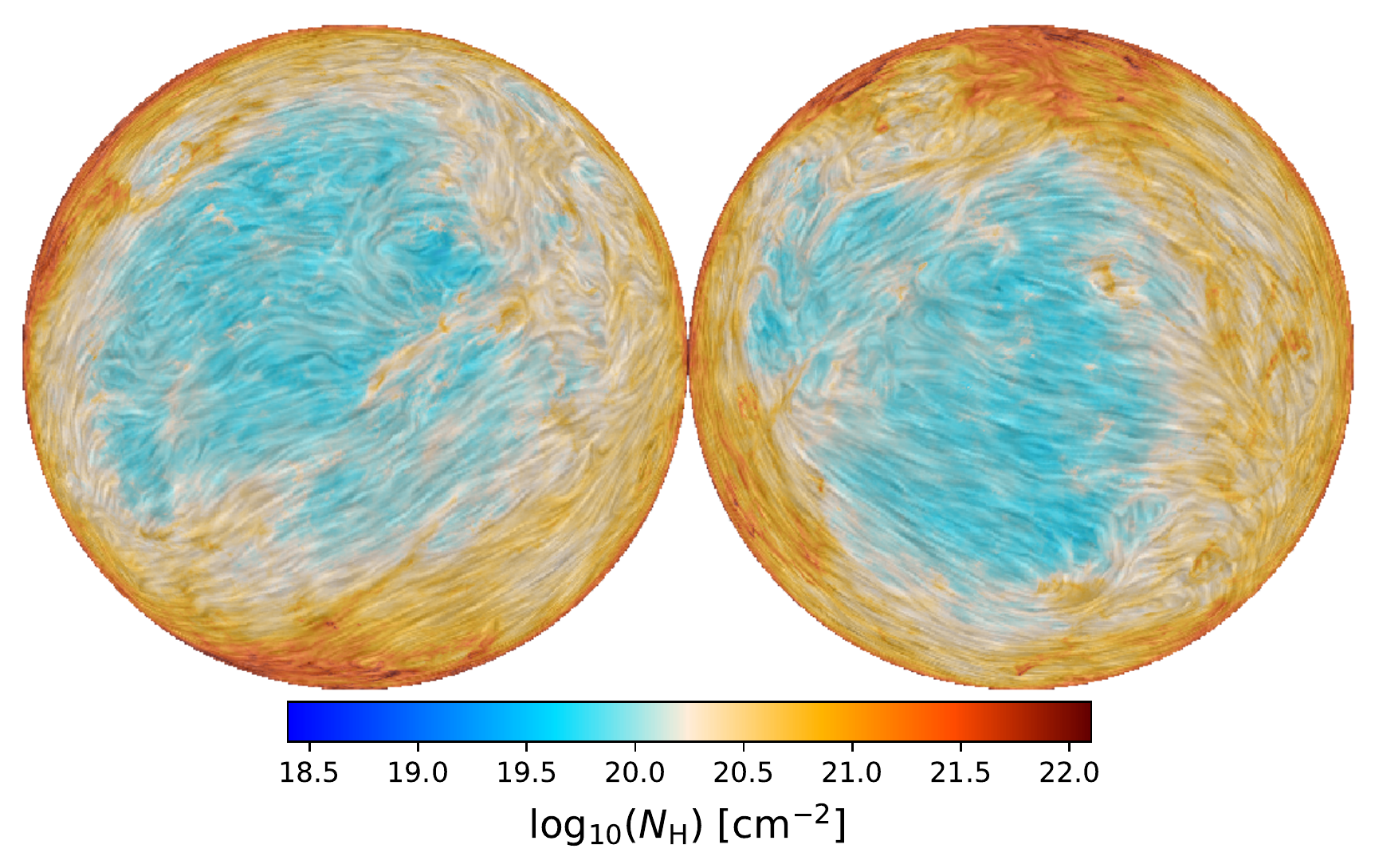}
    \end{subfigure}
    \caption{The same as Fig.~\ref{fig:maps-NH+LIC-lowRes} but for actual observational 353\,GHz dust emission and polarization data of the {\it Planck} mission.}
    \label{fig:planckMaps}
\end{figure*}

In Fig.~\ref{fig:planckMaps}, we present the equivalent of Fig.~\ref{fig:maps-NH+LIC-lowRes} but using the data from the {\it Planck} satellite \citep[][]{PlanckI2016}.
A comparison between our numerical simulation with observational data at $|b|$\,$\lesssim$\,$15^\circ$ makes little sense because we are only able to trace the Galactic midplane within our simulation domain, that is, out to a distance of 250\,pc.
At low- and intermediate Galactic latitudes, the comparison is more meaningful, as the dust exponential scale heights in the Milky Way are around 100\,pc and 90\,pc for the cold and warm components, respectively \citep[see e.g.][]{Misiriotis2006}. 
At these latitudes, we note that our magnetic field lines show a morphological pattern that is qualitatively in agreement with the {\it Planck} data at 353\,GHz.
In our synthetic observations, the homogeneous magnetic field results from the initial conditions in the simulation, which is an initial magnetic field aligned along the $x$ direction.
In the case of {\it Planck}, it is the product of the Galactic magnetic field, which is most likely parallel to the Galactic plane, integrated along the whole LOS, although there are a few regions in the {\it Planck} map where the observed field orientation seems to be produced by nearby clouds, even at low Galactic latitudes, as, for example, toward the Pipe Nebula \cite[][]{alves2007,soler2016}. 

Toward the Galactic caps, the reconstructed magnetic field lines exhibit high turbulence and non-homogeneity, contrasting with the {\it Planck} observations and the previously mentioned theoretical models. 
This fact is confirmed by the differences in angles between S0 and S0uniB and S0justLB and S0uniB, as shown in the middle and bottom panels of Fig.~\ref{fig:polAngleDifference-lowRes}.
These significant differences are particularly prominent towards the southern Galactic cap, which can be explained by the fact that the simulated Local Bubble is open in that direction.
The northern cap also exhibits substantial angle differences, especially in the $y$ direction, that is, orthogonal to the initial mean field orientation.
Most likely, the non-regularity and non-homogeneity of the polarization angle toward the poles are caused by the bubble expansion, which is responsible for the ejection of material from the disk and the possible break-up of the cavity toward the poles, with the consequence that the magnetic field lines become tangled.

Our results indicate that variations in the field direction introduced by either turbulence in the shell or the fragmentation of its walls are significant in explaining the morphological pattern obtained for our Local Bubble candidate. These effects are typically not included in the flux-freezing theoretical models \citep[see e.g.][]{FerriereEtAl1991}.
However, our MHD simulations do not incorporate a global toroidal magnetic field component, which is typically present in rotationally supported spiral galaxies \citep[][]{beck2013,BorlaffEtAl2021}.
This leads to a dichotomy: either the Local Bubble is singularly homogeneous in comparison to the bubble produced in our simulations, or the homogeneous field orientation observed toward the poles in the {\it Planck} polarization data is a result of the combined effects of the dust and magnetic field present behind the Local Bubble.

Further investigation is required to determine whether the plane-of-the-sky field homogeneity at high Galactic latitudes can be attributed to the field in the Local Bubble shell or a coherent large-scale field component is necessary. 
Resolving these issues requires the analysis of numerical simulations that produce realistic Local Bubble analogs embedded in a highly-resolved full disk galaxy. 
These are not available yet.

\subsection{Open vs. closed cap}

SN explosions generate expanding bubbles that can  break out of the Galactic disk and subsequently expand into the halo. 
This is one of the known processes that can cause the Galactic fountain flows \citep[see e.g.][]{Bregman1980, Grand2019, Bish2019}. 
Observational evidence suggests that the Local Bubble shell has been blown out of the disk above and below the Galactic plane, and could be a galactic chimney \citep[][]{Welsh1999}.
Our Local Bubble candidate has an open and a closed cap, which will break in turn in about 10\,Myr, as observed from the analysis of the subsequent snapshots.
We would like to know if it is possible to distinguish between the two configurations through polarization observations.
 
\begin{figure*}
    \centering
    \includegraphics[width=1.0\textwidth]{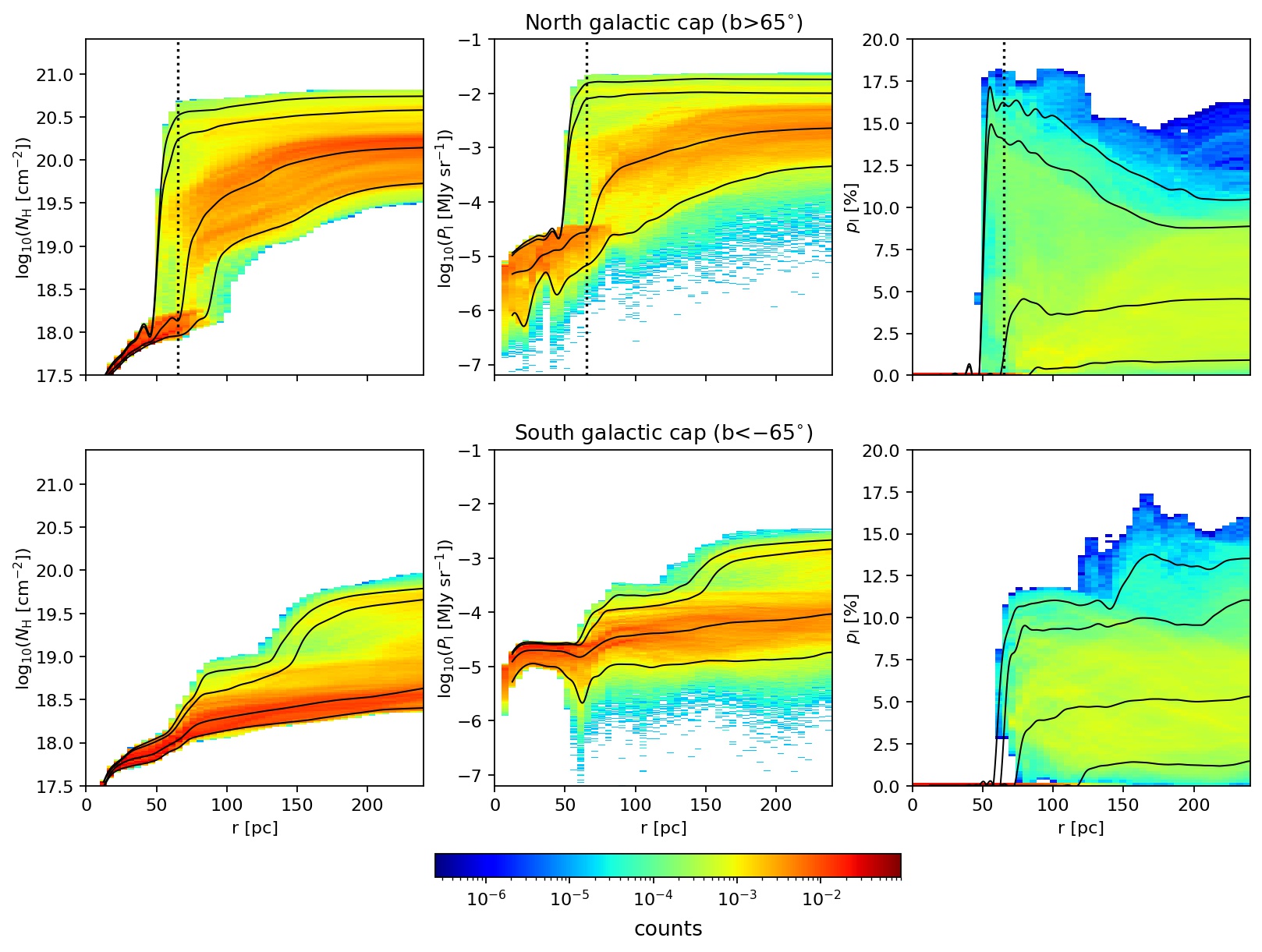}
    \caption{Same as Fig.~\ref{fig:2Dhist_XYplane-lowRes}, but for the northern (closed) and southern (open) Galactic caps, defined for the galactic latitudes $b$\,$>$\,$+65^\circ$ and $b$\,$<$\,$-65^\circ$, respectively.}
    \label{fig:2Dhist_north+southCaps-lowRes}
\end{figure*}

In Fig.~\ref{fig:2Dhist_north+southCaps-lowRes} we report the relative normalized 2D-Histograms regarding $N_{\rm H}$, $P_{\mathrm{l}}$, and $p_{\mathrm{l}}$. 
As expected, the open cap presents low column density and low polarized flux values compared to the closed one. 
This is due to the fact that in the south direction, the gas has been swept from the simulation domain, and the radiation encounters, during its path toward the observer, just some rarefied and isolated clouds of gas that are still inside the bubble; see, for example, the top panel of Fig.~\ref{fig:allCuts-lowRes}.
Furthermore, $N_{\rm H}$ and $P_{\mathrm{l}}$ in the case of an open cap do not undergo a sharp change in their values as in the case of a closed one.

From the orthographic projection in the top panel of Fig.~\ref{fig:maps-NH+LIC-lowRes}, the magnetic field morphology toward the caps can be appreciated. 
We note that it would be difficult to determine  which one is the open and the closed cap, just from the field morphology alone.

\subsection{Caveats}

To produce our synthetic observations, we choose a simulation that follows the dynamical evolution of the complex multi-phase ISM in a 500\,$\times$\,500\,$\times$\,500\,pc$^{3}$ volume.
The  physical properties of the environment, such as the gas density, gas temperature, and magnetic field in our simulation are comparable to the ones of the solar neighborhood, although cosmic rays are missing.
The resolution of our simulation (512 cells per side) allows us to resolve the structure of the ISM in the walls of the Local Bubble candidate, but it limits our study to a portion of the ISM and not to an entire galaxy.

To study the influence of the bubble on the polarization data received by an observer inside it, we isolate the shell of the cavity. However, we acknowledge that our method may not accurately identify the bubble walls in certain directions due to the complexity of gas distribution. Nevertheless, we evaluate the robustness of our results by making slight manual adjustments to the thickness of the bubble walls. We confirm that these modifications do not significantly affect our results.

In our simulation, we have SN explosions which are responsible for blowing cavities that structure the ISM similarly to ours and for ejecting material from the disc, although it is possible that we are underestimating the amount of ejected gas and dust, or that the ejected material is lost from the simulation domain.
The ideal numerical setup for this study should include the large-scale field resulting from the Galactic dynamics and the numerical resolution to detail the structure of the Local Bubble candidates.
However, such a numerical experiment is challenging for the current computational facilities.

\section{Conclusions}\label{sec:conclusions}

We performed an analysis of synthetic dust polarization observations for an observer in the center of a simulated SN-blown cavity with properties similar to the Local Bubble around the Sun. We found that the Local Bubble walls make a non-negligible contribution to the polarized radiation at 353\,GHz detected inside the cavity. This stresses the importance of studying the Local Bubble in detail since all electromagnetic information from objects beyond the cavity must be observed through this local veil of material. 

Specifically, we compared our synthetic dust polarization maps with data from the {\it Planck} satellite \citep[][]{PlanckI2016}. For this, we distinguished three distinct regions on the sky: the disk midplane ($|b| < 15^{\circ}$), low- to intermediate latitudes ($15^{\circ} \le |b| \le 65^{\circ}$), and the Galactic polar regions ($|b| > 65^{\circ}$). 

Because the numerical simulation we used to extract the cavity only allows us to trace the Galactic midplane out to a distance of $250\,$pc \citep{GirichidisEtAl2018b, girichidis2021}, comparing to observational data at $|b| \lesssim 15^\circ$ makes little sense since the full disk is missing. The analysis of this region on the sky will need to be the topic of future studies in which Local Bubble analogues are extracted from high-resolution simulations of the full Galactic disk with star formation, stellar feedback, and a realistic multi-phase description of the ISM. However, we note that already our current model is able to reproduce the observed convergence points of the field at $l \approx \pm 90^{\circ}$.  

At low and intermediate Galactic latitudes ($15^{\circ} \le |b| \le 65^{\circ}$) we found that the magnetic field lines, inferred from dust synthetic polarization observations, display an overall  morphological pattern and polarization degree in good agreement with current models of the magnetic field structure in the Local Bubble walls \citep[see e.g.][]{alves2018} and with the {\it Planck} data at 353\,GHz \citep[][]{PlanckI2016}.

Observations of the polarization  around the north and south Galactic poles ($|b| > 65^{\circ}$) indicate a highly regular and coherent magnetic field structure \citep{PlanckXLIV2016}. Our simulations are in tension with these results, as we found a high degree of small-scale structures, caused by the local turbulence in the expanding bubble walls, as well as indications of open field lines, consistent with the fact that one of the caps of the studied cavity has already burst open into the low-density gas above the disk midplane, and the other cap is expected to do so in a few Myr. As our local ISM model has no global toroidal magnetic field component, as is characteristic for rotationally supported spiral galaxies \citep{beck2013,BorlaffEtAl2021}, we conclude that a coherent large-scale field component may be necessary for explaining the observed polarization pattern at high Galactic latitudes. Again, this emphasizes the need for conducting high-resolution global disk simulations to investigate this further. An alternative explanation could be that our Local Bubble candidate has an unrealistically high level of turbulence and fragmentation. It is therefore important to further study the gas dynamical properties of the Local Bubble and better measure the local velocity dispersion as function of distance from the Sun in different directions. 

Finally, we note that given the current estimates of structure and spatial extent of the Local Bubble  \cite[e.g.][]{alves2018,PelgrimsEtAl2020, ZuckerEtAl2022} it seems likely that the cavity is in the process of bursting into the low-density gas above the disk midplane, or has already done so. Our results indicate that it is difficult to distinguish between an open and a closed Galactic cap based only on  the field morphology derived from polarization measurements. Further data are required, ideally, from the combination of dust extinction measurements and starlight polarization observations to obtain an estimate of the  three-dimensional magnetic field structure of our nearby Galactic environment. 

\section*{Acknowledgements}

We thank the referee for the insightful comments that helped to improve this paper.
We also thank the following people who helped with their encouragement and conversation: Fran\c cois Boulanger, Andrea Bracco, Alyssa Goodman, Jo\~ao Alves, Catherine Zucker, and Theo O'Neill.
Part of the crucial discussions that led to this work took part under the program Milky-Way-Gaia of the PSI2 project funded by the IDEX Paris-Saclay, ANR-11-IDEX-0003-02.
EM joined the research group in Heidelberg thanks to the European Erasmus+ Traineeship programme.
JDS, PG, and RKS acknowledge funding from the European Research Council (ERC) via the Synergy Grant ``ECOGAL -- Understanding our Galactic ecosystem: From the disk of the Milky Way to the formation sites of stars and planets'' (project ID 855130).
RSK and SR acknowledge funding from the Deutsche Forschungsgemeinschaft (DFG) via the Collaborative Research Center (SFB 881, Project-ID 138713538) ``The Milky Way System'' (subprojects A1, B1, B2 and B8) and from the Heidelberg Cluster of Excellence (EXC 2181 - 390900948) ``STRUCTURES: A unifying approach to emergent phenomena in the physical world, mathematics, and complex data'', funded by the German Excellence Strategy. 
RSK also thanks the German Ministry for Economic Affairs and Climate Action for funding in the project ``MAINN -- Machine learning in astronomy: understanding the physics of stellar birth with invertible neural networks'' (funding ID 50OO2206). MCS acknowledges financial support from the Royal Society (URF\textbackslash R1\textbackslash 221118).
The group in Heidelberg is grateful for computing resources and data storage provided through bwHPC and SDS@hd by the Ministry of Science, Research and the Arts of the State of Baden-W\"{u}rttemberg and DFG (grants INST 35/1134-1 FUGG and  35/1314-1 FUGG).

\section*{Data Availability}
The Local Bubble candidate simulation data are publicly available at \url{http://silcc.mpa-garching.mpg.de}. The radiative transfer code {\tt POLARIS} is available at \url{https://portia.astrophysik.uni-kiel.de/polaris/}. The synthetic observations maps and the analysis scripts will be shared upon request.

\newpage
\bibliographystyle{mnras}
\bibliography{references} 



\clearpage
\appendix

\section{Simulating a Local Bubble}\label{Appendix:sims}

\begin{figure*}
    \centering
    \includegraphics[width=0.9\textwidth,angle=0,origin=c]{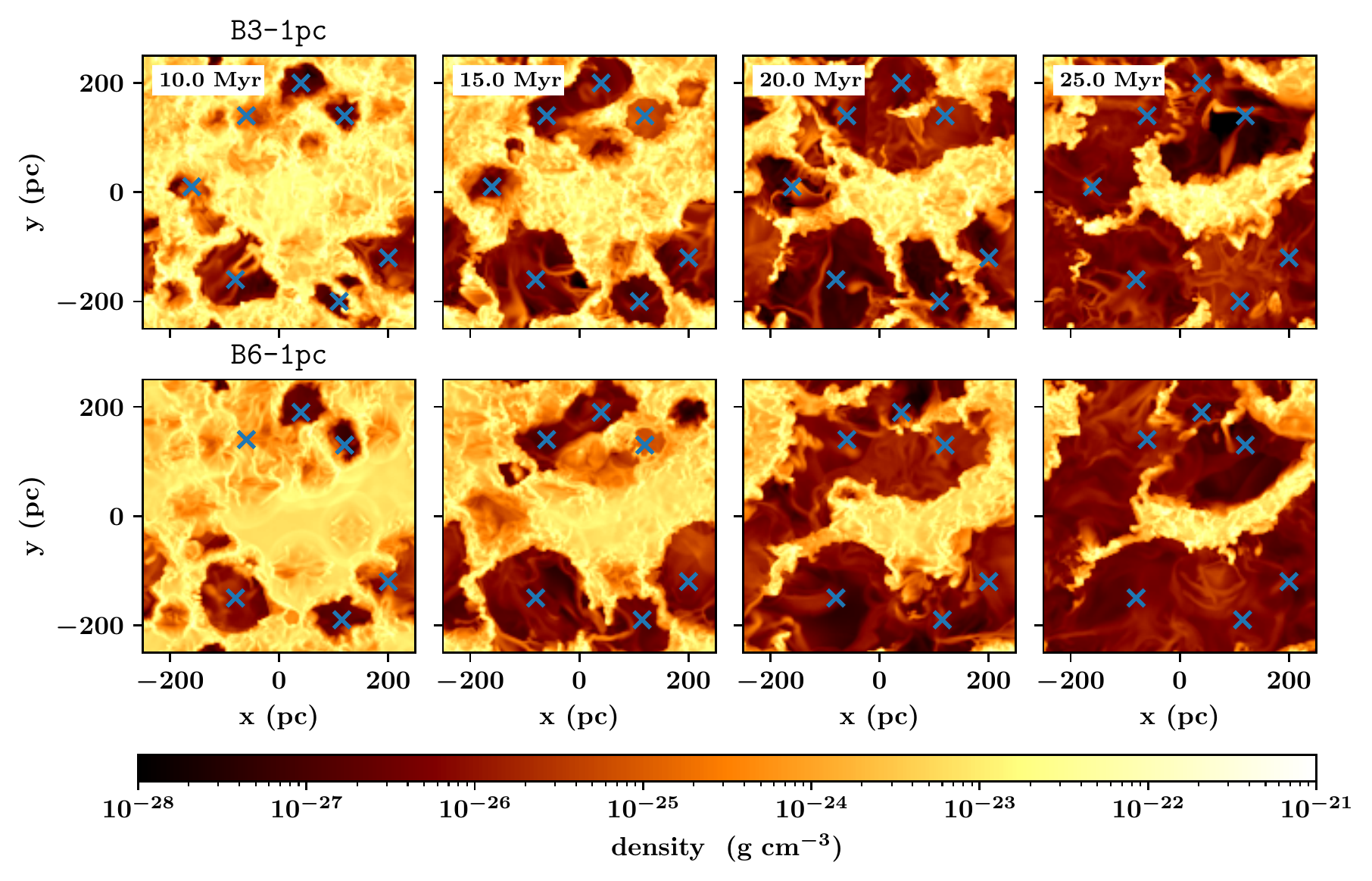}
    \caption{Density at the midplane for two different simulations, B3-1pc (top) and B6-1pc (bottom) for different simulation times (from left to right). The intitial candidate positions for the bubble search are shown in blue crosses. Only at early times we satisfy the constraints for the number of SNe that drives the bubble ($\sim15$).}
    \label{fig:density-time-evol}
\end{figure*}

\begin{figure}
    \centering
    \includegraphics[width=8cm]{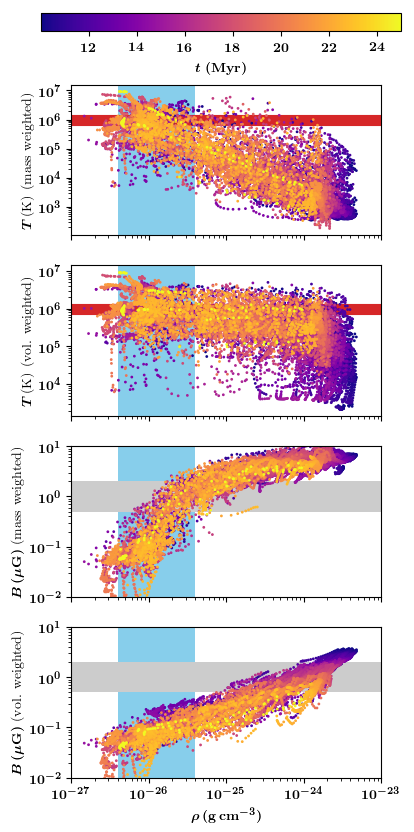}
    \caption{All bubbles from the B3-1pc simulation. Every tenth snapshot from 10 to 39 Myr is included. Shaded areas are approximate observational constraints. Colour-coded is the simulation time. In the early stages, the clouds are denser and colder (dark blue points) compared to later times. If we consider the volume-weighted magnetic field, no bubble falls into the desired range of observational values. However, we only miss the range by less than a factor of 1.5. The mass-weighted counterparts fit very well.}
    \label{fig:all-bubbles-B3}
\end{figure}

\subsection{Details of the numerical setup}

We use the numerical simulations from \cite{GirichidisEtAl2018b} and \cite{girichidis2021}, which are part of the SILCC (SImulating the Life-Cycle of molecular Clouds) project \citep{WalchEtAl2015,GirichidisEtAl2016b}. The setup covers a domain of $500\times500\times500\,\mathrm{pc}^3$ in size. Initially, the gas is uniformly distributed in the $xy$ plane with a vertical stratification in $z$. The gas is initially at rest. 
We analyze two simulations, hereafter referred to as B3-1pc and B6-1pc, with an initial magnetic field set along the $x$ direction and with a central intensity of $B(z=0)$\,$=$\,3\,$\mu$G and $B(z=0)$\,$=$\,6\,$\mu$G, respectively. The magnetic field strength scales in the vertical direction as ${ B(z) = B(z=0)(\rho(z)/\rho(z=0))^{1/2} }$.
The three-phase medium including the filaments, clouds, and voids is the result of SN driving.
The SN explosions rate is constant and based on the Kennicutt-Schmidt (KS) relation \citep{Schmidt1959,Kennicutt1998} for star formation.
Assuming the initial mass function (IMF) presented in \cite{chabrier2003}, the star formation rate surface density from the KS relation translates into an SN rate of $15\,\mathrm{Myr}^{-1}$ for the simulation box, where every star above eight solar masses explodes as an SN.

The SN explosions in the selected simulation set are a mixture of 20\% type~Ia and 80\% type~II.
The type~Ia SN explosions have a uniform distribution in $x$ and $y$ and a normal distribution in $z$ with a scale height of $300\,\mathrm{pc}$.
The type~II SN explosions are split between individual SN explosions and clustered ones.
The individual SN explosions, which correspond to 60\% of the type~II, are placed randomly in the $x-y$ plane and normally distributed in $z$ with a scale height of $50\,\mathrm{pc}$.
The remaining 40\% of the type~II are clustered assuming clusters with a lifetime of $40\,\mathrm{Myr}$ and a total number of SN explosions, $N$, ranging between 7 and 20,  randomly drawn from a distribution with a probability scaling with $N^{-2}$.

The time evolution of the gas distribution is shown in Fig.~\ref{fig:density-time-evol} for B3-1pc (top panels) and B6-1pc (bottom panels) with some identified bubbles marked with blue crosses. The most prominent and largest bubbles correspond to clustered SN feedback.

\subsection{Identification of the Local Bubble in the simulations}

Initially, we select a set of cavities by eye in each simulation and identified their central position $\bm{p}_{0}$\,$=$\,$[x_{0},y_{0},z_{0}]$.
For each candidate, we investigate spheres around $\bm{p}_{0}$ with radius $r$ between 40 and 70\,pc.
We compute the average density, the total mass in ionized gas, and the mass- and volume-weighted magnetic field and temperature. 

Among all the spheres for each candidate, we select the ones which satisfy the following conditions

\begin{itemize}
    \item $10^{-27}\,\mathrm{g\,cm}^{-3} \le \rho \le 4\times10^{-26}\,\mathrm{g\,cm}^{-3}$;
    \item $T \geq 5\times10^5\,\mathrm{K}$;
    \item $0.5\,\mu\mathrm{G} \le B \le 2\,\mu\mathrm{G}$.
\end{itemize}

For the magnetic field strength we use a selection criterion which corresponds nicely to the values measured by \cite{XuHan2019}. The magnetic field volume weighted quantities are below 1\,$\mu$G. The mass weighted field directly reflects the field in the shell of the bubble, since the shell mass dominates over the mass in the void. In B3-1pc the mass weighted field of the bubble as a function of radius quickly exceeds the value of 2\,$\mu$G \citep[upper limit in][]{XuHan2019}, as a result of local compression with respect to the initial field value. For this reason we decided to use the B3-1pc instead of the B6-1pc simulation.

A scatter plot with the selected bubbles is depicted in Fig.~\ref{fig:all-bubbles-B3} showing from top to bottom the mass and volume-weighted temperature as well as magnetic field strength as a function of average cloud radius. The observational constraints are depicted in shaded colour. The colour coding of the points corresponds to the simulation time.

\section{Dust heating, grain alignment and synthetic polarization maps with {\tt POLARIS}}\label{Appendix:POLARIS}

{\tt POLARIS} is a three-dimensional Monte-Carlo (MC) dust continuum
radiative transfer code dedicated to post-process MHD simulations \citep{Reissl2016}. The code, making use of the physical quantities (e.g. density, temperature, velocity, and magnetic fields) provided by MHD simulations together with an arbitrary number of radiation sources, can compute among other things the dust temperature, the grain alignment efficiency and produce synthetic multi-wavelength intensity and polarization maps.

Linear and circular polarization arise from non-spherical dust grains; these elongated grains have the tendency to align their minor axis with the magnetic field direction and to emit light preferentially along their major. It is therefore necessary to define a specific dust model beforehand e.g. resembling the conditions in the ISM (see Sect. \ref{subSect:post-processing}).
The corresponding cross sections, for sub-mm emission, can be computed as
 
\begin{equation}
    C_{\rm{x}} = \int^{a_{\rm{max}}}_{a_{\rm{min}}} N_{\mathrm{d}}(a) \big[ C_{\perp}(a) + C_{\parallel}(a) \big] \, \mathrm{d}a,
    \label{eq:CrossSection}
\end{equation}
and 
\begin{equation}
    \Delta C_{\rm{x}} \cong \sin^2 \theta \int^{a_{\rm{max}}}_{a_{\rm{alg}}} R(a) N_{\mathrm{d}}(a) \times \big[ C_{\perp}(a) - C_{\parallel}(a) \big] \, \mathrm{d}a.
    \label{eq:DeltaCrossSection}
\end{equation}
Here, the quantity $C_{\rm{X}}$ may represent the cross sections of extinction, scattering, and absorption whereas $a_{\rm{alg}}$ is the minimal radius of grain alignment, $\theta$ is the angle between the magnetic field direction and the LOS, $R(a)$ is the Rayleigh reduction factor accounting for the grain alignment efficiency, $C_{\perp}(a)$ and $C_{\parallel}(a)$, respectively,  are the pre-calculated cross sections perpendicular and parallel to the grain minor principal axis \citep[see][]{Reissl2016}.
{\tt POLARIS} calculates then the radiation field based on an MC approach in multiple wavelengths $\lambda$. Once the radiation field is defined,  the dust temperature $T_{\mathrm{d}}$ is determined assuming an equilibrium between the absorption of photons, and  thermal dust emission \citep{Lucy1999, Bjorkman2001, Reissl2016} as well as the grain alignment efficiency.

For the computation of the grain alignment efficiency, a key aspect in modelling synthetic dust polarization, {\tt POLARIS} follows the physics of the radiative torque alignment theory \citep[RAT;][]{Lazarian2007}. The hallmark of this theory it is that anisotropic radiation spins-up irregular grains which subsequently align, in the case of paramagnetic grains, with the magnetic field direction; this effect is opposed by the random gas particles bombardment. The spin-up process can be quantified by the ratio \citep[][]{Lazarian2007, Hoang2013}

\begin{equation}
    \Big(\frac{J_{\rm{RAT}}}{J_{\rm{g}}} \Big)^2 \propto a \Big(\frac{1}{n_{\rm{g}}} T_{\rm{g}} \int \lambda Q_{\rm{RAT}} \gamma_{\lambda} u_{\lambda} d\lambda \Big)^2,
\end{equation}
where $J_{\rm{RAT}}$ and $J_{\rm{g}}$ are the dust grain angular momentum induced, respectively, by radiation and by gas collisions.
Based on the properties of the MC calculated radiation field {\tt POLARIS} determines the anisotropy factor $\gamma_{\lambda}$ and the energy density $u_{\lambda}$ \citep[for a detailed description we refer to][]{Reissl2020}. For the grain alignment efficiency, $Q_{\rm{RAT}}$, the following parametrization is used

\begin{equation}
    \label{eqn:Q-RAT}
  Q_{\rm{RAT}} = \cos \Psi 
    \begin{cases}
      0.4 & \text{if $\lambda\leq 1.8 \, a$}\\
      0.4 \times \Big(\frac{\lambda}{1.8 \, a}\Big)^{-2.6} & \text{otherwise}\\
    \end{cases}       
\end{equation}
which is based on the average of a large ensemble of different grain shapes \citep{Herranen2019}. In eq.\eqref{eqn:Q-RAT}, $\Psi$ is the angle between the anisotropic radiation field and the magnetic field direction \citep[for details see][]{Lazarian2007, Reissl2016, Reissl2020}.

Dust grains with a stable alignment are the ones with the ratio $J_{\rm{RAT}}/J_{\rm{g}} > 3$ \citep[see e.g.][for details]{Hoang2013}; it is then possible to compute a characteristic grain size $a_{\rm{alg}}$, defined as the size where $J_{\rm{RAT}}/J_{\rm{g}} = 3$, above which all paramagnetic grains ($a>a_{\rm{alg}}$) i.e. will contribute to the dust polarization $R(a)=1$. Otherwise, for grain sizes $a<a_{\rm{alg}}$ the Rayleigh reduction factor is ${ R(a)=0 }$. Furthermore, {\tt POLARIS} allows only silicate grains to align but graphite grains cannot due to their different paramagnetic properties \citep[see e.g.][]{Hunt1995, Draine1996, Hoang2014-b, andersson2015}. Hence, $R(a)=0$ for all considered graphite grain sizes.

Once the dust temperature and the grain alignment efficiency are computed, {\tt POLARIS} can be used to simulate synthetic multi-wavelength intensity and polarization maps. To accomplish this, the code solves the radiative transfer equation in all four Stokes parameters $\textbf{S} $ simultaneously. This problem, in the most general form, can be expressed as \citep[see e.g.][]{Martin1971,Jones1979}
\begin{equation}
    \frac{d}{\mathrm{d}\ell}\textbf{S} = - \hat{K} \textbf{S} + \textbf{J},
    \label{eq:RT}
\end{equation}
where $\textbf{J}$ the emissivity and $\hat{K}$ the 4$\times$4 Müller's matrix describing the extinction, absorption, and emission are fully defined by the cross sections defined in Eq. \ref{eq:CrossSection} and, Eq. \ref{eq:DeltaCrossSection}, respectively. Consequently, the emitted total and linearly polarized intensity $I$ and $P_{\mathrm{l}}$, respectively,  along each LOS accumulate as
\begin{equation}
    \mathrm{d}I \propto n_{\rm{g}} C_{\rm{abs}} B_{\lambda}(T_{\rm{d}}) \mathrm{d}\ell,
\end{equation}
and 
\begin{equation}
    \mathrm{d}\ell P_{\mathrm{l}} \propto n_{\rm{g}} \Delta C_{\rm{abs}} \times B_{\lambda}(T_{\rm{d}}) \mathrm{d}\ell,
\end{equation}
where $n_{\rm{g}}$ is the gas number density, $B_{\lambda}(T_{\rm{d}})$ is the Planck function.

 {\tt POLARIS} solves Eq.\ref{eq:RT} along the LOS for each path element $\mathrm{d}\ell$ by utilizing a Runge-Kutta solver \citep[see e.g.][]{Ober2015, Reissl2019}. Moreover, the code allows the definition of plane detectors, for an observer placed outside the grid, or of {\tt HEALPix} detectors for an observer placed inside the simulation. For more technical details about {\tt POLARIS}, we again refer the reader to \cite{Reissl2016}, \cite{Reissl2019}, and \cite{Reissl2020}, respectively.


\bsp	
\label{lastpage}
\end{document}